\documentclass[pre,
aps,
amsmath,
amssymb,
twocolumn,
superscriptaddress,
longbibliography
]{revtex4-2}

\usepackage{ragged2e}    \usepackage{marginnote}   
\usepackage{cases}
\usepackage[makeroom]{cancel}
\usepackage[export]{adjustbox}
\usepackage{hyperref} 
\usepackage{caption,subcaption}
\usepackage{enumerate}
\usepackage{xr-hyper}
\usepackage{xcolor}
\usepackage{xr}

\begin{document}
\title{Analytical theory of coherent radiation and radiation friction in laser--plasma collisions}
\author{E.~G. Gelfer}\email{egelfer@gmail.com}
\affiliation{ELI Beamlines facility, The Extreme Light Infrastructure ERIC,
Dolni Brezany 252 41, Czech Republic}
\author{A.~M.~Fedotov}
\affiliation{National Research Nuclear University MEPhI, Moscow, 115409,  Russia}
\affiliation{
Institute of Applied Physics of the Russian Academy of Sciences, Nizhny Novgorod 603950, Russia}
\author{M.~P.~Malakhov}
\affiliation{National Research Nuclear University MEPhI, Moscow, 115409, Russia}
\affiliation{Skolkovo Institute of Science and Technology, 
Skolkovo, 121205, Russia}
\author{Th.~Benahmed}\altaffiliation[Current address: ]{Deutsches Elektronen--Synchrotron DESY, Notkestr. 85, 22607 Hamburg, Germany}
\affiliation{ELI Beamlines facility, The Extreme Light Infrastructure ERIC,
Dolni Brezany 252 41, Czech Republic}
\author{J.~Custodio}
\affiliation{ELI Beamlines facility, The Extreme Light Infrastructure ERIC,
Dolni Brezany 252 41, Czech Republic}
\author{O.~Klimo}
\affiliation{ELI Beamlines facility, The Extreme Light Infrastructure ERIC,
Dolni Brezany 252 41, Czech Republic}
\affiliation{FNSPE, Czech Technical University in Prague, Prague, Czech Republic}
\author{S.~Weber}
\affiliation{ELI Beamlines facility, The Extreme Light Infrastructure ERIC,
Dolni Brezany 252 41, Czech Republic}

\begin{abstract}
    We develop an analytical theory of coherent (scaled quadratically with the number of particles) radiation and coherent radiation friction in a head–on collision of a dense charged particle bunch with an intense laser pulse. We demonstrate that the low–frequency coherent radiation in the forward and backward directions dominates the energy-momentum losses of a mildly relativistic bunch and can result in a substantial enhancement of the overall radiation friction as compared to the incoherent case. We derive the scaling laws for the average momentum losses of the bunch over the collision with respect to laser intensity, pulse duration, and particle bunch parameters, and show their robustness with respect to laser polarization and the shape of the particle distribution in the bunch.
\end{abstract}

\maketitle

\section{Introduction}
When charged particles collide with a laser pulse, they emit electromagnetic radiation losing energy and momentum. From classical viewpoint, this effect, conventionally called radiation reaction or radiation friction (RF) \cite{dirac_prs1938,landau2,jackson_book},  can be formally described by introducing a RF force $\mathbf{F}_{RF}$ into the equation of motion
\begin{equation}\label{eqm}
\frac{d\mathbf{p}}{dt}=\mathbf{F}_L+\mathbf{F}_{RF},
\end{equation}
where $\mathbf{F}_L=e(\mathbf{E}+[\mathbf{v}\times\mathbf{B}]/c)$ is the Lorentz force, $\mathbf{E}$ and $\mathbf{B}$ are the electric and magnetic fields, $e$, $\mathbf{p}$ and $\mathbf{v}$ are the charge, momentum and velocity of the particle. 

RF has been recently  observed experimentally in laser--particles collisions \cite{cole_prx2018,poder_prx2018,los_natcomm2026} and it is foreseeable that RF can be important in practical applications of strong laser--plasma interaction, such as electron and ion acceleration \cite{voronin_jetp1965,zeldovich_ufn1975,dipiazza_lmp2008,tamburini_njp2010,tamburini_nimpra2011,tamburini_pre2012,chen_ppcf2010,capdessus_pre2015,gelfer_njp2021}, generation of waves in a rarefied plasma \cite{gelfer_scirep2018, gelfer_ppcf2018}, or strong magnetic fields in a dense plasma  \cite{liseykina_njp2016,popruzhenko_njp2019}. It also modifies the spectrum of radiation itself \cite{koga_pop2005,dipiazza_prl2010,neitz_prl2013,thomas_prx2012}. 

Assuming $B\sim E$, classical description is valid as long as the quantum parameter $\chi\sim\frac{\gamma E}{E_{cr}}$, which measures the ratio of the electric field in the particle rest frame to the critical field of quantum electrodynamics (QED) $E_{cr}=m^2 c^3/e\hbar$  \cite{sauter1931,schwinger1951}, is small \cite{ritus1985,dipiazza_rmp2012}. Here $m$ and $\gamma$ are the particle mass and Lorentz factor. In what follows, we assume the classical regime $\chi\ll1$ of radiation and radiation friction.

 Several expressions for the RF force have been proposed in the literature \cite{dirac_prs1938,landau2,eliezer_prs1948,ford_pla1991}, see also the review \cite{burton_contphys2014}. The one introduced by Landau and Lifshitz (LL) \cite{landau2}, is now commonly accepted as a convenient approximation for the description of RF in the classical regime \cite{gonoskov_rmp2022,fedotov_physrep2023}.  

It is conventionally believed that RF is relevant in interactions of ultraintense lasers with ultrarelativistic particles, see the reviews \cite{dipiazza_rmp2012,burton_contphys2014,gonoskov_rmp2022,fedotov_physrep2023}. Indeed, according to the estimates  \cite{dipiazza_lmp2008,koga_pop2005} based on the LL approach, RF is substantial for 
\begin{equation}\label{LLcond}
\mathcal{R}=\mu a_0^2\gamma\omega_L T\gtrsim1,\quad \mu=\frac{4\pi}{3}\frac{r_e}{\lambda_L},
\end{equation}
where $a_0=e E_0/m\omega_L c$ is the amplitude of the dimensionless field strength, $\omega_L$, $\lambda_L$ and $T$ are the frequency, wavelength and duration of the laser pulse, and $r_e=e^2/mc^2\approx 2.8\cdot 10^{-13}$~cm is the classical electron radius. For an optical petawatt class laser $\mu\sim 10^{-8}$, $\omega_LT\sim 10-100$, so that Eq.~\eqref{LLcond} can be satisfied only for $a_0\gg1$ and/or $\gamma\gg 1$.

However, Eq.~(\ref{eqm}) and the condition Eq.~(\ref{LLcond}) have been derived under the assumption that the particles emit radiation independently and  incoherently. This assumption might be not fulfilled for a dense particle bunch  \cite{hartemann_pre2000,gonoskov_pre2015,gelfer_prr2024,gelfer_mre2024}. Since the energy of the emitted coherent radiation scales with the number of radiating particles $N$ as $N^2$ \cite{schwinger1945,michel_prl1982,hirschmugl_pra1991,hartemann_pre2000,gonoskov_pre2015,vieira_natphys2021,malaca_natphot2024,gelfer_prr2024,gelfer_mre2024,quin_ppcf2025,quin_prr2025}, the coherence can strongly enhance RF. 

Here we investigate coherently enhanced RF in a head-on collision of a charged particle bunch with a plane wave laser pulse. We provide a detailed derivation of the analytical theory of the effect and demonstrate its robustness with respect to the models in use for the laser pulse and the particle bunch.

In the companion paper \cite{prl_arxiv} we summarize the main physical results, provide their numerical validation with three-dimensional particle--in--cell (PIC) simulations, confront the coherently enhanced RF with the ordinary incoherent one and discuss the prospects for experimental observation of coherent RF in such a setup.

The paper is organized as follows. In Sec.~\ref{sec_rad} we derive an analytical approximation for the radiation spectrum of a bunch of charged particles. To this end we follow the approach of Ref.~\cite{gelfer_prr2024} but, whereas there the main focus was on the high frequency part of the radiation spectrum, here we mostly consider radiation at small frequencies, which contributes most to the energy and momentum transfer in the regime of coherently enhanced RF. In Sec.~ \ref{sec_mom} we derive an amount of momentum lost on average by a particle in a dense bunch. First we discuss the known result for a single particle, then calculate the momentum losses assuming the coherent radiation is dominant. In Sec.~\ref{sec_disc} we discuss the obtained analytical results with an emphasis on their robustness with respect to the laser polarization and the bunch shape. Section~\ref{sec_sum} summarizes our findings. The Appendix supports the discussion of Sec.~\ref{sec_disc} by providing a detailed calculation of RF for an alternative bunch shape.

\section{Radiation of a particle bunch colliding with laser pulse}\label{sec_rad}

We start with a general formula for the radiation emitted by a bunch of $N$ particles colliding with a laser pulse. Assuming the waist $w$ of the laser pulse is much larger than both the laser wavelength and the radius of the bunch, $w\gg\lambda_L,\,R$ we consider the laser pulse as a plane wave pulse propagating along $x$ axis, for which the vector potential $\mathbf{A}(\phi)$ depends solely on its phase $\phi=\omega_L(t-x/c)$ [$\omega_L$ is the laser frequency]. Then it is convenient to parametrize particle trajectories and momenta by the phases $\phi_j=\omega_L(t-x_j/c)$, where $j$ ($1\leq j\leq N$) labels the particles. In these terms the angular--frequency distribution of the radiation can be written in the form \cite{jackson_book}
\begin{equation}\label{dE}
\begin{split}
&\frac{d\mathcal{E}}{d\omega d\Omega}=\frac{\omega^2}{4\pi^2\omega_L^2c}\times\\
&\left|\sum\limits_{j=1}^N \frac{e_j}{p_{-,\,j}}\int[\mathbf{n}\times[\mathbf{n}\times\mathbf{p}_j(\phi_j)]]e^{i\omega\varkappa(\phi_j)}d\phi_j\right|^2,
\end{split}
\end{equation}
where $\varkappa(\phi_j)=\phi_j/\omega_L+(x_j(\phi_j)-\mathbf{n}\mathbf{r}_j(\phi_j))/c$, $\mathbf{n}$ and $\omega$ are the direction and frequency of the emitted radiation, $e_j$, $\mathbf{r}_j(\phi_j)$ and $\mathbf{p}_j(\phi)$ are the charge, coordinates and momenta of the $j$-th particle of the bunch, and $p_-=\gamma m c-p_x$ is the lightfront momentum conserved along the trajectory in a plane wave field.

\subsection{Initial phase of a particle in a bunch}\label{app_phase}
Consider the trajectories of particles colliding head-on with a plane wave laser pulse. Assuming the same initial momenta of all the particles, it follows from the equation of motion without RF that the phase dependence of the particle momenta is the same for all particles
\begin{equation}\label{pj}
\mathbf{p}_j(\phi_j)\equiv\mathbf{p}(\phi_j).
\end{equation}

Next, from $d\mathbf{r}_j/d\phi_j=c\mathbf{p}/\omega_Lp_- $ we obtain
\begin{equation}\label{rj}    \mathbf{r}_j=\mathbf{r}_j(\hat\phi)+\mathbf{r}(\phi_j),\quad \mathbf{r}(\phi_j)=\int\limits_{\hat\phi}^{\phi_j}\frac{c\mathbf{p}(\phi')}{\omega_L p_-}d\phi',
\end{equation}
where $\hat{\phi}$ is a (common for all the particles) initial value of the phase. 

By substituting Eqs.~(\ref{pj}) and (\ref{rj}) into Eq.~(\ref{dE}) we find that Eq.~(\ref{dE}) can be factorized \cite{schiff_rsi1946,hartemann_pre2000,gelfer_prr2024}
\begin{equation}\label{C}
\frac{d\mathcal{E}}{d\omega d\Omega}=\mathcal{C}\frac{d\mathcal{E}^{(1)}}{d\omega d\Omega}, \quad \mathcal{C}=\left|\sum_{j=1}^Ne^{i\Phi_j}\right|^2,
\end{equation}
where $d\mathcal{E}^{(1)}/d\omega d\Omega$ is the single particle radiation spectrum corresponding to Eq.~(\ref{dE}) with $N=1$ and $\Phi_j=\omega_L(x_j(\hat\phi)-\mathbf{n}\mathbf{r}_j(\hat\phi))/c$.

For practical reasons it is more convenient to express the phase shifts $\Phi_j$ of the particles through their initial positions  $\mathbf{r}_j^0\equiv\mathbf{r}_j(\phi_j(t_0))$ at a given common time $t_0$ rather than $\mathbf{r}_j(\hat\phi)$ at a common phase. Let us choose the initial time $t_0$ such that all the particles are still outside the pulse moving freely with the initial momentum $p_x=-p_0$, i.e. $x_j(\phi)$=$x_j(\hat\phi)-p_0(\phi-\hat{\phi})/\omega_Lp_-$. Then up to a common constant term we have
\begin{equation}\label{xphi} x_j(\hat\phi)=x_j^0\left(1-\frac{p_0}{p_-}\right)+\frac{p_0t_0}{p_-},\quad \mathbf{r}_{\bot,j}(\hat\phi)=\mathbf{r}^0_j.
\end{equation}
Finally, taking into account that in the relativistic case $p_0\approx p_-/2$, from Eq.~(\ref{xphi}) we arrive at 
\begin{equation}\label{phij}
\Phi_j=\omega[(1-\cos\theta)x_j^0/2
-\mathbf{n}_{\bot} \mathbf{r}_{j,\bot}^0]/c.
\end{equation}
Here $x_j^0$, $\mathbf{r}^0_{j,\bot}$ are the particle initial longitudinal and transverse coordinates.

Note that the factor $1/2$ in the longitudinal contribution was missed in previous works \cite{hartemann_pre2000,gelfer_prr2024}. It appears due to that the laser pulse propagates toward the particles and does not appear in a static field, e.g. for synchrotron radiation  \cite{schiff_rsi1946}.

\subsection{Coherent and incoherent contributions}
Consider now factor $\mathcal{C}$ that relates single and multiple particle spectra in Eq.~(\ref{C}). Representing $\mathcal{C}=\sum_{j,k}e^{i(\Phi_j-\Phi_k)}$ and decomposing the sum to $N$ diagonal and $N(N-1)$ off-diagonal terms \cite{schiff_rsi1946} we get
\begin{equation}\label{C1}
\mathcal{C}=N(1-\alpha)+N^2\alpha,    
\end{equation}
where $\alpha=|\left<e^{i\Phi}\right>|^2$ is the modulus squared average of the particle initial phase factor. The terms $N(1-\alpha)$ and $N^2\alpha$ correspond to the incoherent and coherent parts of the radiation, respectively. The second (coherent) term is dominant for $\alpha N\gg1$. In what follows we will call $\alpha$ the coherence factor.

For a given initial density distribution $n(\mathbf{r^0})$ of particles in a bunch the coherence factor $\alpha$ can be calculated as
\begin{equation}\label{alpha}
    \alpha=\left|\frac{1}{N}\int n(\mathbf{r^0}) e^{i\Phi(\mathbf{r^0})}d\mathbf{r}^0\right|^2,
\end{equation}
where $N=\int n(\mathbf{r^0}) d\mathbf{r}^0$. In particular, for a Gaussian density distribution $n(x,\mathbf{r}_\bot)=n_0\exp\left(-\frac{x^2}{(L/2)^2}-\frac{r_\bot^2}{R^2}\right)$ employed in Ref.~\cite{prl_arxiv} we have 
\begin{equation}\label{alphag}
\alpha=e^{-\frac{\omega^2}{8c^2}\left(L^2\sin^4\frac{\theta}{2}+4R^2\sin^2\theta\right)}.
\end{equation}

\begin{figure*}[t]
\centering
\includegraphics[width=0.45\textwidth]{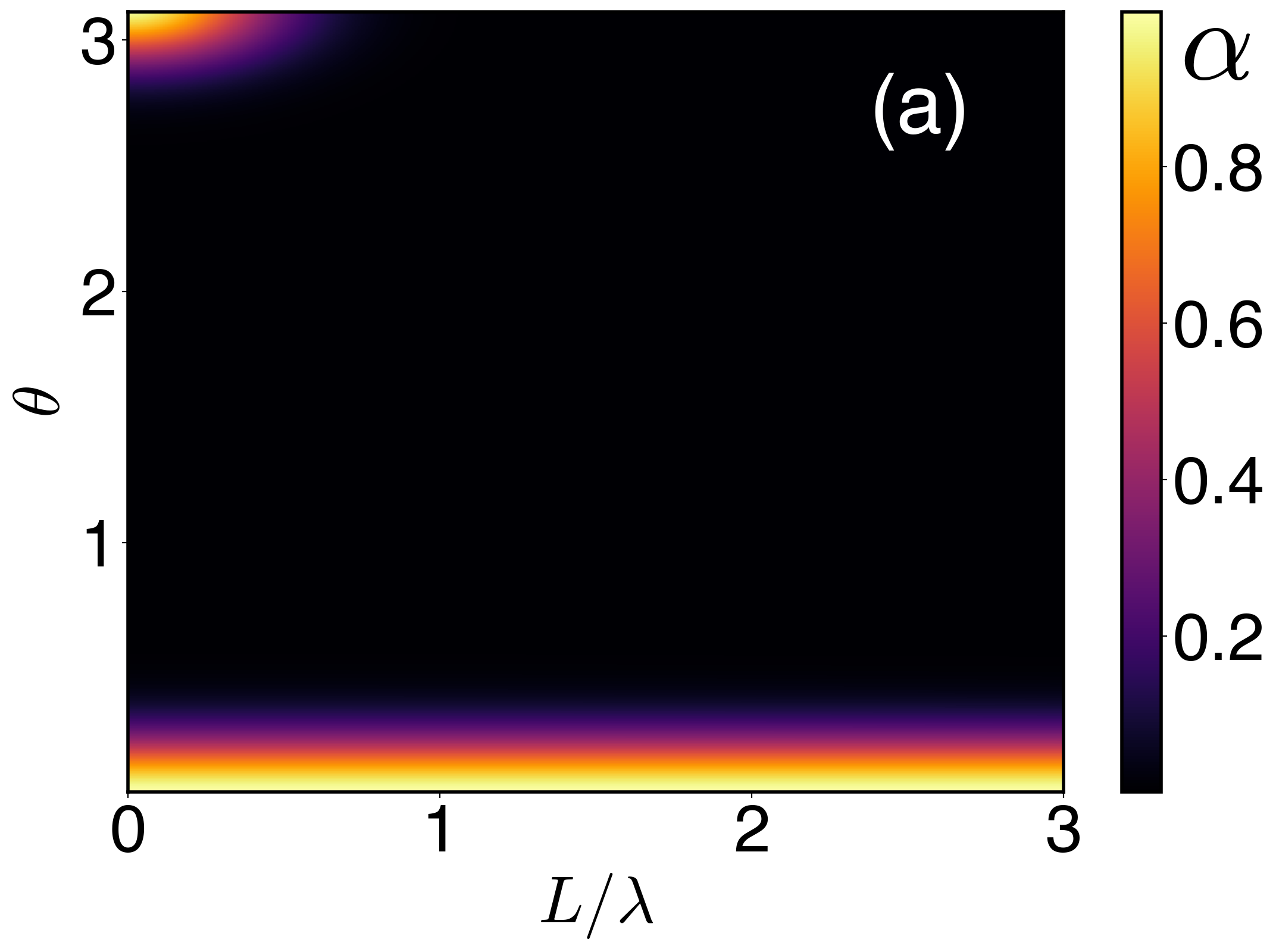}
\includegraphics[width=0.45\textwidth]{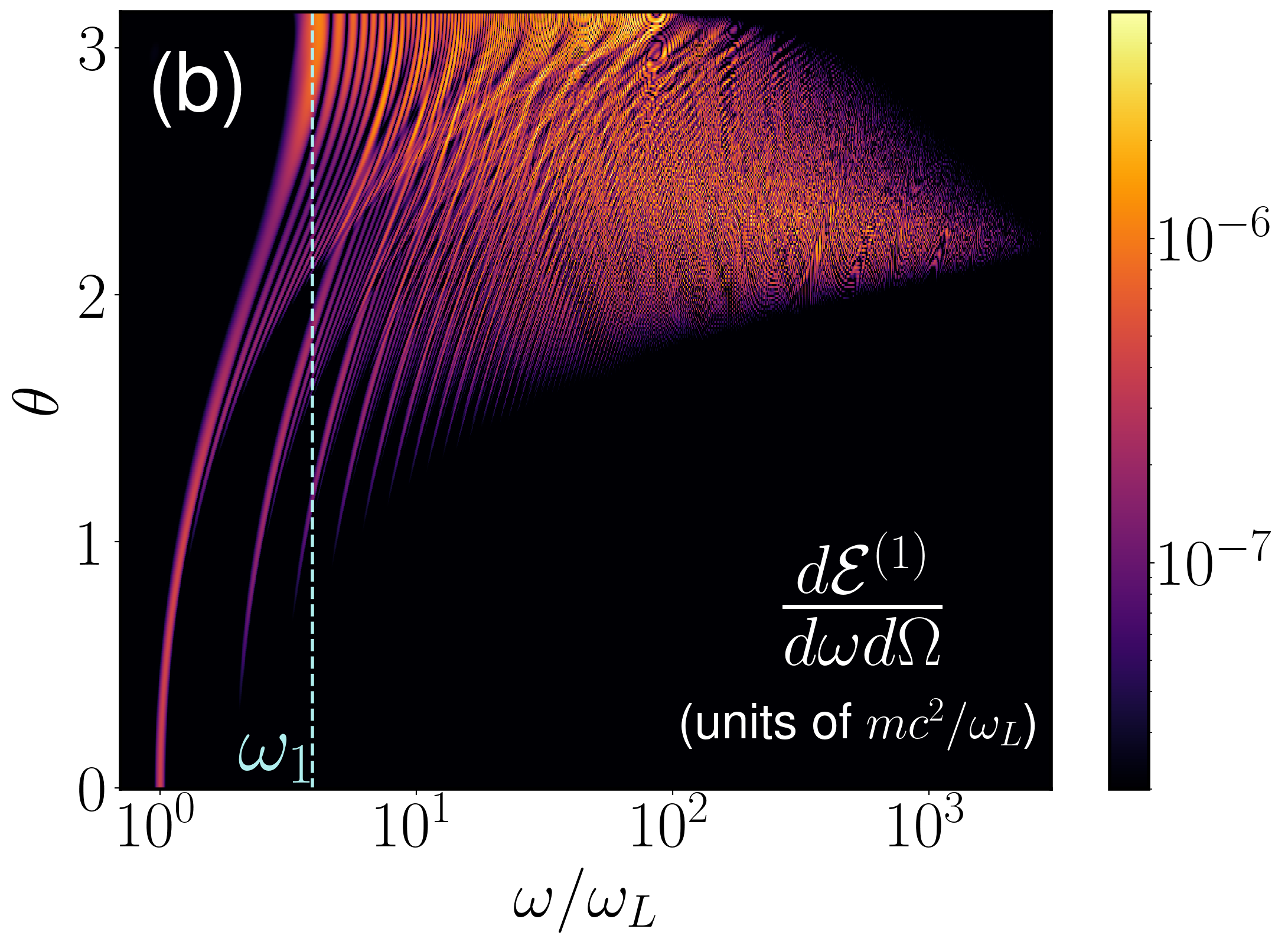}
\caption{Coherence factor (a) and radiation spectrum of a single particle (b) for a head-on collision of electrons with a plane wave laser pulse; $\omega_1$, is the fundamental harmonic of the backward radiation defined in Eq.~(\ref{om1}); $R=\lambda$, $p_0=5mc$, $T\approx 50$~fs.
}\label{fig_rad}
\end{figure*}

It is worth noting that $\alpha\approx1$ for a small bunch such that $R,L\ll \lambda$, where $\lambda=2\pi c/\omega$ is the wavelength of the emitted radiation. This means that the radiation is fully coherent in all directions and $N$ particles of charge $e$ can be replaced with a single particle of charge $Ne$. If $R$ and/or $L$ $\gtrsim\lambda$, then $\alpha$ has a global maximum $\alpha=1$ at $\theta=0$ (forward radiation) and, depending on $R$ and $L$, an additional peak at $\theta=\pi$ (backward radiation). It is easy to see that in the Gaussian case this additional peak appears if $L<2\sqrt{2}R$. However, for $L\gg\lambda$ its contribution is strongly suppressed, see Fig.~\ref{fig_rad}~(a). 

\subsection{Single particle radiation in forward and backward directions}\label{app_single}
According to Eqs.~(\ref{C}) and (\ref{C1}), the $N$-particle spectrum is a product of the coherence factor (\ref{alpha}) and the single particle spectrum $d\mathcal{E}^{(1)}/d\omega d\Omega$. While the coherence factor can be calculated analytically at least for simple shapes of the particle bunch, the single particle spectrum in a finite laser pulse in general has to be evaluated numerically \cite{thomas_prstab2010,boca_pra2009,seipt_pra2011}, see the result of such evaluation in Fig.~\ref{fig_rad}~(b).

However, as shown above, see also Ref.~\cite{prl_arxiv}, for a large bunch (of size larger than the emitted wavelength at least in one dimension), the coherent radiation narrowly peaks around the backward and/or forward directions. Therefore, for analytical evaluation of the coherent radiation spectrum, it is enough approximate the single particle spectrum in these directions, where it can be cast to the form \cite{hartemann_pre1996,gelfer_prr2024}
\begin{equation}\label{dE1}
\left.\frac{d\mathcal{E}^{(1)}}{d\omega d\Omega}\right|_{\theta=0,\pi}=\frac{e^4\omega^2}{4\pi^2c\omega_L^2p_-^2}\left|\int\mathbf{A}(\phi) e^{\frac{i\omega}{\omega_L}(\phi+\upsilon(\phi))}d\phi\right|^2,
\end{equation}
where $\upsilon(\phi)=0$ for $\theta=0$  (forward radiation) and $\upsilon(\phi)=2\omega_Lx(\phi)/c$ for $\theta=\pi$ (backward radiation). Here 
$$x(\phi)=\frac{c}{\omega_L}\left[\frac{p_\parallel}{p_-}\phi+\frac{m^2c^2}{2p_-^2}\int\limits_{-\infty}^\phi \hat{\mathbf{A}}^2(\psi)d\psi\right]$$ 
is the longitudinal coordinate of the particle and $\hat{\mathbf{A}}(\phi)=e\mathbf{A}(\phi)/mc$ is the dimensionless vector potential.

In particular, the forward radiation spectrum is proportional to the spectrum of the incident laser pulse. For an incoming circularly polarized pulse of duration $T$ with a Gaussian temporal envelope 
\begin{equation}\label{A}
\mathbf{A}=\frac{mca_0}{e}e^{-\phi^2/(\omega_LT)^2}\{0,\cos\phi,\sin\phi\},
\end{equation}
the forward scattered spectrum
\begin{equation}\label{dE1f}
\left.\frac{d\mathcal{E}^{(1)}}{d\omega d\Omega}\right|_{\theta=0}\approx \frac{e^2\omega^2T^2}{8\pi c}\frac{m^2a_0^2c^2}{p_-^2}e^{-\frac{(\omega-\omega_L)^2T^2}{2}},
\end{equation}
is also Gaussian, centered at the laser carrier frequency $\omega_L$. 

For the back scattered radiation we have
\begin{equation}\label{backint}
\begin{split}
\left.\frac{d\mathcal{E}^{(1)}}{d\omega d\Omega}\right|_{\theta=\pi}&=\frac{e^2 p_-^2\zeta^2}{4\pi^2m^2c^3}\left|\int\hat{\mathbf{A}}(\phi) e^{if(\phi)}d\phi\right|^2,\\
&f(\phi)=\zeta\left(\phi+\int\limits_{-\infty}^\phi \hat{\mathbf{A}}^2(\psi) d\psi\right),
\end{split}
\end{equation}
where $\zeta=(mc/p_-)^2\omega/\omega_L$. The integral in Eq.~\eqref{backint} can be evaluated by the stationary phase approximation \cite{seipt_lasphys2013,kharin_pra2016,gelfer_prr2024}. 

Representing Eq.~(\ref{A}) in terms of complex exponentials, we observe that for a long pulse $\omega_LT\gg 1$ each of the phases $if(\phi)\pm i\phi$ in Eq.~(\ref{backint}) has two stationary points 
\begin{equation}\label{phi1}
\phi^1_\pm=\pm\omega_LT\sqrt{\frac{1}{2}\ln\frac{\zeta a_0^2}{1-\zeta}},
\end{equation}
and
\begin{equation}\label{phi2}
\phi^2_\pm=\pm\omega_LT\sqrt{\frac{1}{2}\ln\frac{\zeta a_0^2}{-1-\zeta}}.
\end{equation}
Since for all real $\zeta$ the points $\phi^2_\pm$ are imaginary and such that $|\mathrm{Im}\phi^2_\pm|>|\mathrm{Im}\phi^1_{\pm}|$, their contribution is  exponentially suppressed and  can be neglected. In contrast, stationary points (\ref{phi1}) are real in the range $1/(1+a_0^2)<\zeta<1$, which corresponds to 
\begin{equation}\label{freqinterval}
\hat\omega=\frac{p_-^2\omega_L}{m^2c^2(1+a_0^2)}<\omega<\frac{p_-^2\omega_L}{m^2c^2}=\omega_*.
\end{equation}
Outside the range \eqref{freqinterval}, in particular for $\omega>\omega_*$, all the saddle points are purely imaginary and the radiation spectrum falls out exponentially. Such high frequencies are, however, irrelevant to the coherent effects of our primary interest here. 

Applying the standard formula for the stationary phase approximation, we arrive at 
\begin{equation}\label{dE1b}
\left.\frac{d\mathcal{E}^{(1)}}{d\omega d\Omega}\right|_{\theta=\pi}\approx \frac{e^2}{4\pi c}\frac{\omega T}{\nu}(1+\sin\eta),\quad \hat\omega<\omega\ll \omega_*,
\end{equation}
where 
\begin{equation}\label{eta}
\begin{split}
&\eta=\frac{\omega_LT}{2\omega_*}\left[2\nu(\omega-\omega_*)+\sqrt{2\pi}a_0^2\omega\,\mathrm{erf}\left(\frac{\nu}{\sqrt{2}}\right)\right],\\
&\nu=\sqrt{2\ln\frac{\omega a_0^2}{\omega_*-\omega}},
\end{split}
\end{equation}
and $\mathrm{erf}(z)=(2/\sqrt{\pi})\int_0^z e^{-\zeta^2}d\zeta$ is the error function. 

\begin{figure}
    \centering    \includegraphics[width=0.95\linewidth]{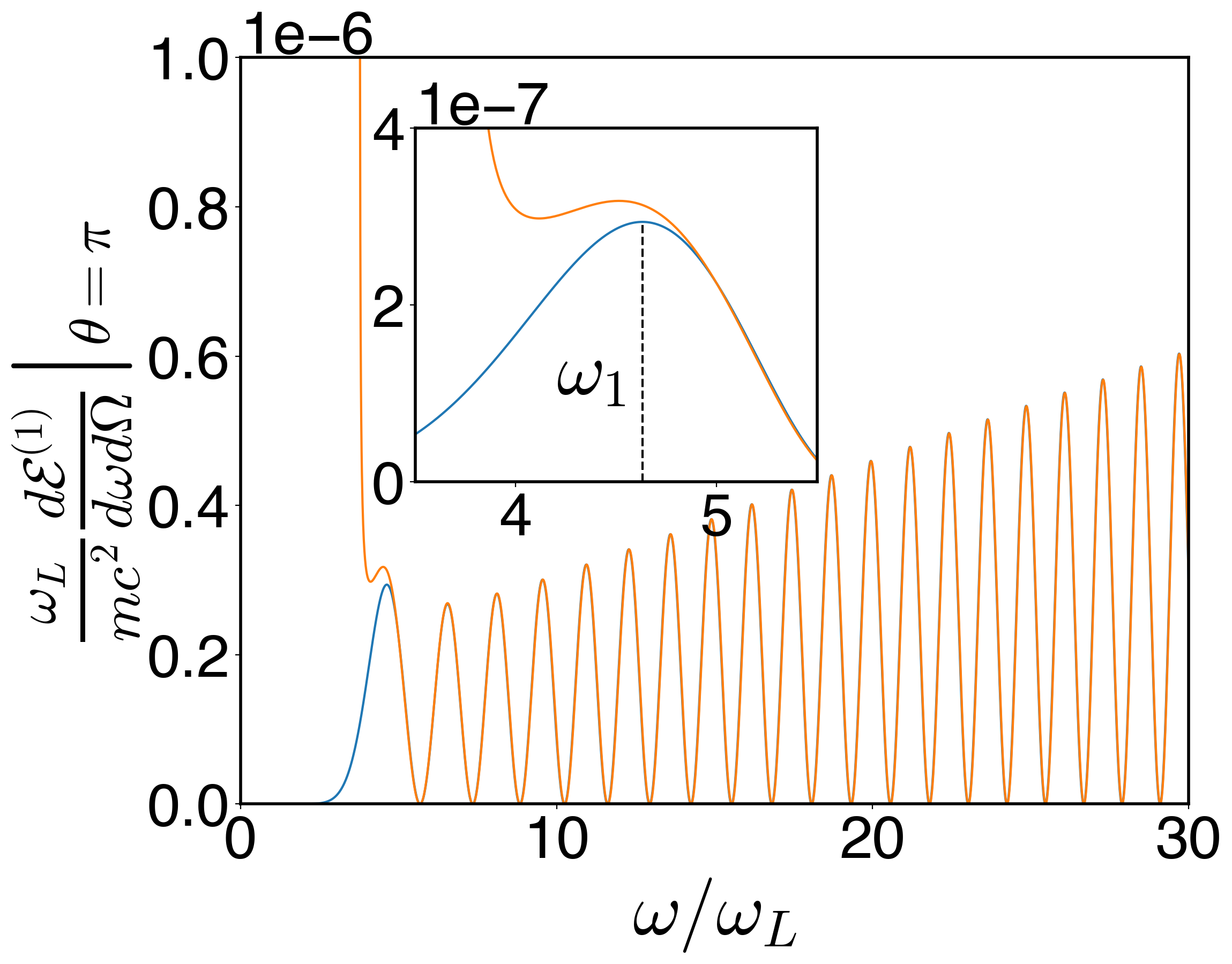}
    \caption{Single particle radiation spectrum in backward direction evaluated numerically (blue) and its analytical approximation Eq.~(\ref{dE1b}) (orange). Other parameters:  $a_0=\gamma_0=5$, $T=67$ fs, $\lambda_L=1\mu$m. Inset: the same plot zoomed in near the first maximum $\omega_1$.}    \label{fig_backspect}
\end{figure}

Approximation (\ref{dE1b}) is compared to the numerically evaluated backward radiation spectrum in Fig.~\ref{fig_backspect}. It is seen that this approximation works reasonably well to the right of $\hat\omega$, where it diverges because the saddle points $\phi^1_+$ and $\phi^1_-$ merge at the origin making the second derivative of the phase vanishing. Nevertheless, Eq.~(\ref{dE1b}) still remains reasonably accurate even at $\omega=\omega_1$, the position of the first maximum of the spectrum (see the inset in Fig.~\ref{fig_backspect}), hence can be used to estimate $\omega_1$. 

To this end we write $\omega_1=\hat\omega(1+\delta)$ assuming $\delta\ll1$ and require $\eta(\omega_1)=\frac{\pi}{2}$, where $\eta(\omega)$ is introduced in Eq.~(\ref{eta}). A straightforward calculation  gives
\begin{equation}\label{om1}
\delta\approx \frac{1}{2}\left(\frac{3\pi}{2\omega_LT}\right)^{2/3}, 
\end{equation}
and it is seen that the implied assumption $\delta\ll1$ indeed holds for $\omega_L T\gg1$.

Finally, from Eq.~\eqref{om1} in the same approximation $\omega_L T\gg 1$ we obtain
\begin{equation}\label{om1f}
\omega_1\approx \omega_L \xi^2\left[1+\frac{1}{2}\left(\frac{3\pi}{2\omega_L T}\right)^{2/3}\right],
\end{equation}
and, in virtue of Eq.~\eqref{dE1b},
\begin{equation}\label{dE1bm}
\left.\frac{d\mathcal{E}^{(1)}}{d\omega d\Omega}\right|_{\theta=\pi, \omega=\omega_1}\approx \frac{e^2}{c}\frac{3\xi^2}{4}\left(\frac{2\omega_L T}{3\pi}\right)^{4/3},
\end{equation}
where $\xi=\frac{p_-}{mc\sqrt{1+a_0^2}}$. 

\subsection{Coherent radiation of a particle bunch}

At this point, we approximate the angular--frequency distribution of the coherent radiation of the particle bunch by
\begin{equation}\label{dEcoh}
    \frac{d\mathcal{E}_{coh}}{d\omega d\Omega}\approx \alpha N^2 \left.\frac{d\mathcal{E}^{(1)}}{d\omega d\Omega}\right|_{\theta=0,\pi},
\end{equation}
where the coherence factor is defined in Eq.~(\ref{alpha}) and we replace the exact single-particle spectrum with the spectrum Eq.~(\ref{dE1f}) of forward scattering for $0\leq\theta<\pi/2$ and with the spectrum Eq.~(\ref{dE1b}) of backward scattering for $\pi/2<\theta\leq\pi$.

\begin{figure*}[t]
\centering
\includegraphics[width=0.45\textwidth]{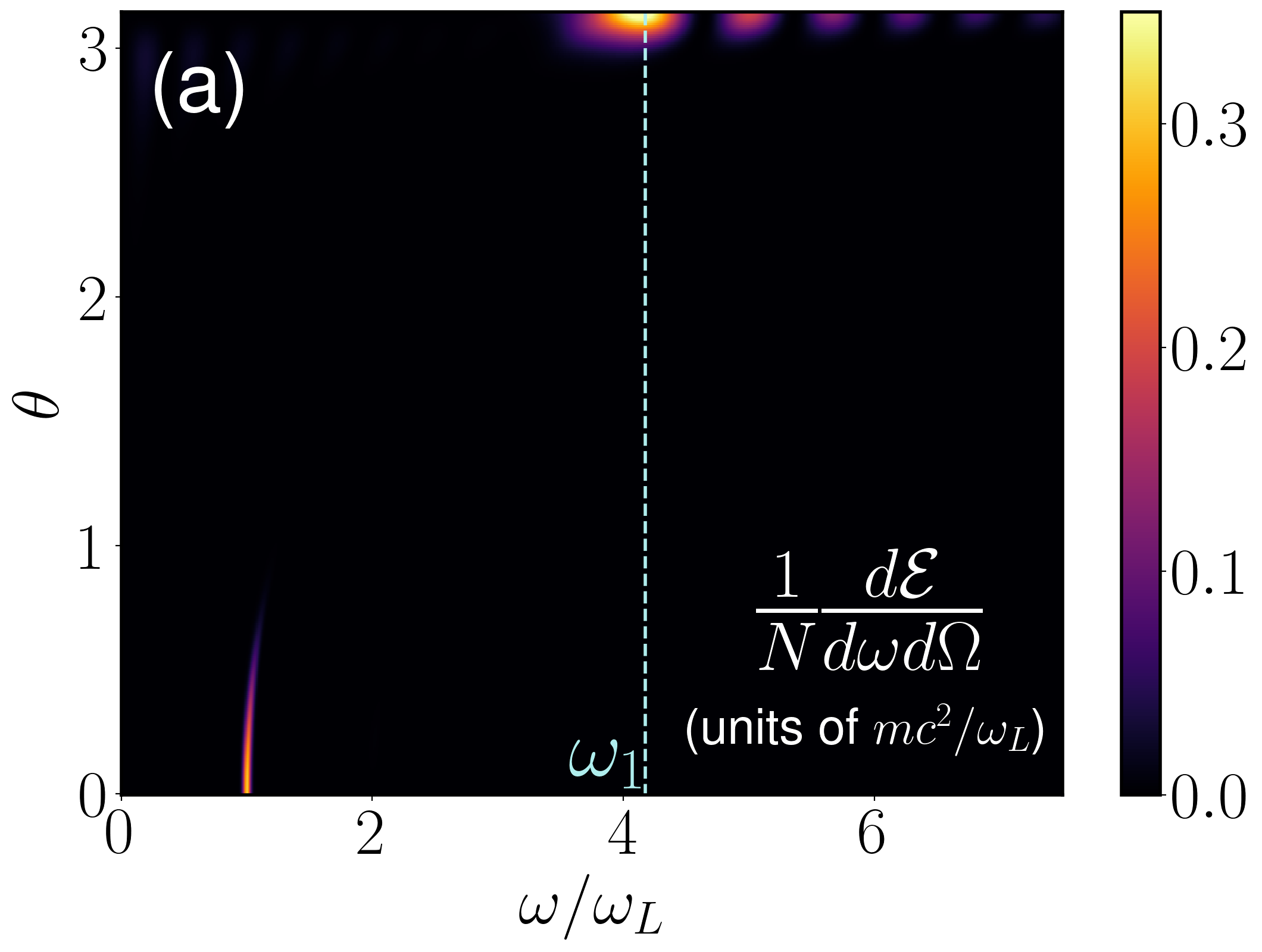}
\includegraphics[width=0.45\textwidth]{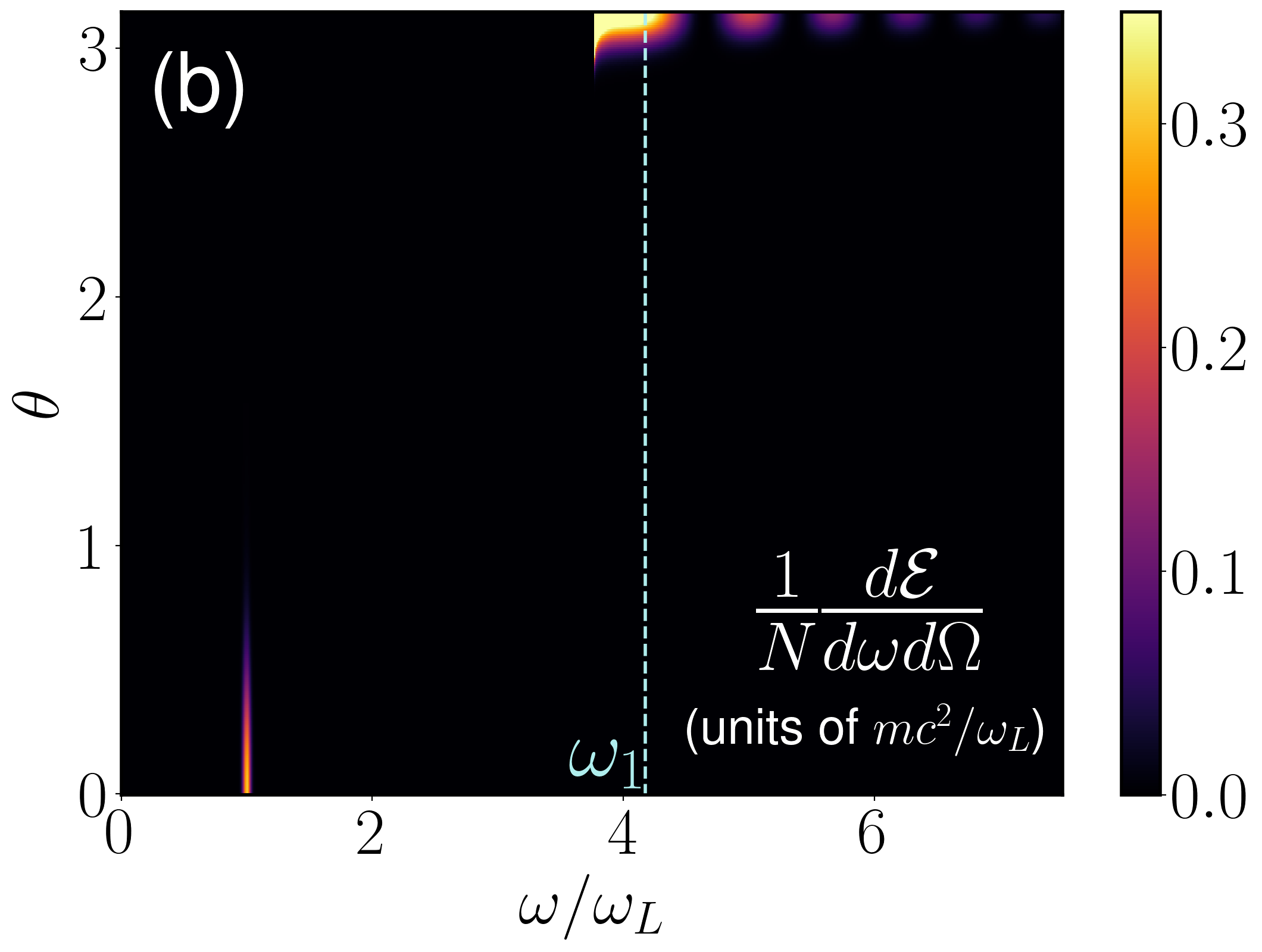}
\caption{
Per-particle radiation spectrum of the Gaussian particle bunch [see Eq.~(\ref{alphag})] of length $L=0.1\lambda_L$, radius $R=0.5\lambda_L$, $n=0.01n_c$, Lorentz factor $\gamma=5$, colliding head on with a plane wave laser pulse with the Gaussian envelope Eq.~(\ref{A}), $a_0=5$, $\omega_LT=15\pi$ (50 fs). (a) -- numerical evaluation of Eq.~(\ref{C}), (b) -- analytical estimate Eq.~(\ref{dEcoh}).}
\label{fig_cohspect}
\end{figure*}

Comparing the approximation Eq.~(\ref{dEcoh}) with Eq.~(\ref{C}), where the single particle spectrum is evaluated numerically, as in Fig.~\ref{fig_rad}~(b), we find that they match very well \footnote{Note that to make the angular distribution wider and better pronounced these plots were made by choosing both $R,L<\lambda_L$. This, however, is worse for the accuracy of our approximation (see below), so that the figure actually corresponds to a worse scenario.}, see Fig.~\ref{fig_cohspect}. As discussed, the coherent radiation is concentrated near the forward and backward scattering directions. In the forward direction, only a narrow frequency band at $\omega\approx\omega_L$ contributes to the spectrum. All this is in a clear correspondence with the structure of the single particle spectrum in the forward direction, see Eq.~(\ref{dE1f}) and Fig.~\ref{fig_rad}~(b). In the backward direction, the largest contribution comes from the fundamental harmonics $\omega_1$, see Eq.~(\ref{om1}).

The approximation Eq.~(\ref{dEcoh}) fails in two major cases. First, for $R,L\ll\lambda$ the coherent radiation is no longer narrowly concentrated in the forward or backward directions, so that the single particle radiation spectrum $d\mathcal{E}^{(1)}/d\omega d\Omega$ has to be found by a fair numerical evaluation of the integral in Eq.~\eqref{dE}. 
Second, the single-particle estimate Eq.~(\ref{dE1b}) for the backscattering spectrum fails for  $\omega<\omega_1$, see Fig.~\ref{fig_backspect}. Since this is not vital for our consideration here, we will present a refined analytical calculation of the single particle spectrum, which resolves this issue, in a forthcoming publication \cite{malakhov_prep}. 

\section{Momentum loss of a single particle and a bunch of particles}\label{sec_mom} 

Now let us turn to RF acting on particles in a laser pulse. As a measure of RF we use the amount of momentum taken away from the particles by radiation. We start with a known case of a single particle colliding with a pulsed plane electromagnetic wave. Then we consider a bunch of particles and express its momentum loss using the light-front momentum conservation in terms of contributions  associated with forward and backward scattered coherent radiation.

\subsection{Single particle}

For a laser pulse  taken as a plane wave solution of the Maxwell equations, such as, e.g., Eq.~(\ref{A}), the equation of motion Eq.~(\ref{eqm}) with the RF term in LL form can be solved exactly \cite{dipiazza_lmp2008}. In particular, for the light component $p_-=\gamma mc-p_\parallel$ of energy-momentum in a circularly polarized pulse, we have
\begin{equation}\label{pmphi}
p_-(\phi)=\frac{p_{-,0}}{1+\mu\frac{p_{-,0}}{mc}\int_{-\infty}^\phi a^2(\phi')d\phi'},
\end{equation}
where $p_{-,0}$ is the initial value of $p_-$, $\boldsymbol{a}(\phi)=e\mathbf{E}(\phi)/m\omega_Lc$ is the dimensionless field strength (note that we direct it along the electric field rather than along the vector potential), $\phi=\omega_L(t-x/c)$ is the phase and $\mu$ was defined in Eq.~(\ref{LLcond}). The exact expression for longitudinal momentum $p_\parallel$ is more involved, see Ref.~\cite{dipiazza_lmp2008}, but in the relativistic case it can be roughly approximated simply by $p_\parallel\approx -p_-/2$. 

Assume that RF is moderately strong, that is $\Delta p_\parallel/p_{0}\ll1$, where $p_0=p_\parallel(-\infty)$ is the initial value of $p_\parallel$. Then we immediately obtain 
\begin{equation}\label{deltapll}
\Delta p_\parallel^{LL}\approx \sqrt{2\pi} \mathcal{R}p_0,
\end{equation}
where $\mathcal{R}$ was defined in Eq.~(\ref{LLcond}). The factor $\sqrt{2\pi}$ comes due to the specific profile of the laser pulse envelope defined in Eq.~(\ref{A}). Note that up to this numerical factor Eq.~(\ref{deltapll}) agrees with the condition Eq.~\ref{LLcond}) for the significance of RF.

The same result also follows from the calculation of the total energy radiated by the particle. Using the expression  $d\mathcal{E}_{tot}^{(1)}/dt=-2r_e/3mc (dp_\mu/d\tau dp^\mu/d\tau)$ for the total radiated power \cite{jackson_book}, where $p^\mu$ is the 4-momentum of the particle and $\tau$ is its proper time, one obtains \cite{seipt_pra2013}
\begin{equation}\label{E1}
\mathcal{E}^{(1)}_{tot}=\mu c\int\limits_{-\infty}^\infty \gamma a^2 p_- d\phi.
\end{equation}
Assuming $\gamma\gg 1$, $\gamma\gg a_0$ and $\Delta p_\parallel\ll p_0$, the quantities $\gamma$ and $p_-$ remain almost constant and can be taken out of the integral. Furthermore, since an ultrarelativistic particle actually radiates almost along its propagation direction \cite{ritus1985}, we have $\Delta p_\parallel\approx \mathcal{E}_{tot}^{(1)}\approx 2\mu mc^2\gamma^2\int a^2(\phi)d\phi$, in agreement with Eq.~(\ref{deltapll}). 

\subsection{Particle bunch}

Next, consider the average momentum loss $\left<\Delta p_\parallel\right>=\Delta P_\parallel/N$ of a particle in a bunch, where $\Delta P_\parallel$ is the total momentum loss of the whole bunch. 
Since it is apriori unknown how the RF force is modified to account for coherent radiation of the bunch, we calculate the momentum taken away by the radiation by means of the energy--momentum conservation using the coherent spectrum derived in Sec.~\ref{sec_rad}. 

It is worth noting that we cannot calculate the energy loss of particles by the integration of the $N$-particle spectrum Eq.~(\ref{C}) over the solid angle and frequency. Indeed, the energy of the emitted radiation is taken not only from the particles but also from the laser pulse. On the other hand, the light cone momentum of the electromagnetic field $P_-^{EM}$ changes from zero (for a plane wave laser pulse) to a positive value after scattering. Its increase is provided solely by the particles and $\Delta P_-=-\Delta P_-^{EM}$, where $P_-$ and $P_-^{EM}$ denote the combination $P_-=\mathcal{E}/c-P_\parallel$ evaluated for the particle bunch and for the electromagnetic field, respectively. Assuming for simplicity that the particles are relativistic both before and after the collision with the laser pulse and therefore $p_\parallel=-p_-/2$, we get~\footnote{Note that the structure of Eq.~\eqref{dP-} resembles the notion of a momentum-transfer cross section that typically arises in considerations of momentum transfer in particle collisions.}
\begin{equation}\label{dP-}
    \left<\Delta p_\parallel\right>\approx\frac{\pi}{cN}\int\limits_0^\infty d\omega\int\limits_0^\pi d\theta \frac{d\mathcal{E}}{d\omega d\Omega}(1-\cos\theta) \sin\theta,
\end{equation}
where, assuming the coherent radiation is dominant, we approximate the $N$-particle spectrum by Eq.~(\ref{dEcoh}). The structure of the approximated coherent radiation spectrum Eq.~(\ref{dEcoh}) suggests calculation of the forward and backward contributions to RF, corresponding to $\int_0^{\pi/2}d\theta$ and $\int_0^{\pi/2}d\theta$ in Eq.~(\ref{dP-}), separately. 

Consider first the forward contribution. For a long pulse $\omega_L T\gg1$, only a narrow band of frequencies close to the laser frequency $\omega_L$ contributes to the integral over $\omega$, which thus casts to 

\begin{equation}\label{dpfg}
    \left<\Delta p^f_\parallel\right>\approx mc\frac{\pi^2}{8\sqrt{2}}\frac{n}{n_c}\frac{a_0^2}{\gamma_0^2}\omega_L T\mathcal{G},
\end{equation}
where $\gamma_0$ is the initial $\gamma$--factor of the particles in the bunch, $n$ and $n_c=m\omega_L^2/4\pi e^2$ are the density of the bunch and the plasma critical density, respectively, and it is taken into account that $N=\pi^{3/2}nR^2L/2$.

The remaining integral over $\theta$ takes the form 
\begin{equation}\label{ggint}
 \mathcal{G}=\frac{\rho}{4\pi}\int\limits_0^{\pi L/2\lambda_L}x e^{-\frac{x^2}{2}+\rho x^2\frac{\lambda_L}{\pi L}-\rho x}dx,
\end{equation}
where $x=(\pi L/\lambda_L)\sin^2(\theta/2)$, $\rho=8\pi R^2/L\lambda_L$. 

The geometric factor $G$ includes the entire dependence of the forward contribution to RF on the sizes $R$ and $L$ of the bunch. To simplify Eq.~(\ref{ggint}) further, consider three cases:
\begin{enumerate}[a)]
    \item $R,L\ll\lambda_L$: 
    the whole argument of the exponential in Eq.~(\ref{ggint}) is small, by replacing the exponential with unity we obtain    \begin{equation}\label{gga}
        \mathcal{G}(R, L\ll\lambda_L)\approx\frac{\pi^2 R^2L}{4\lambda_L^3};
    \end{equation}    
    \item $\rho\gg1$ and $R\gtrsim\lambda_L$: in this case the effectively contributing $x\lesssim \rho^{-1}\ll1$ and hence $\rho x^2\lambda_L/(\pi L)\lesssim (\lambda_L/R)^2/(8\pi^2)\ll1$. Therefore we can retain just the last term in the exponential in Eq.~(\ref{ggint}), thus arriving at 
\begin{equation}\label{ggb}
\mathcal{G}\left(\frac{R^2}{L\lambda_L}\gg1, R\gtrsim\lambda_L\right)\approx\frac{L\lambda_L}{32\pi^2R^2};
\end{equation}
    \item $\rho\ll1$ and $L\gtrsim\lambda_L$: here the effectively contributing $x\lesssim 1$ and we can retain just the first term in the exponential to obtain 
\begin{equation}\label{ggc}
\mathcal{G}\left(\frac{R^2}{L\lambda_L}\ll1,L\gtrsim\lambda_L\right)\approx\frac{2R^2}{L\lambda_L}.
\end{equation}
\end{enumerate}

In case Eq.~(\ref{gga}) the whole bunch emits as a single macroparticle, hence the coherence of the scattering gets maximal and the forward contribution to the momentum loss is naturally proportional to the number of particles in the bunch $N\sim nR^2L$. Note, however, that in this case the radiation angular distribution is no longer narrow as assumed and hence the overall accuracy of Eq.~(\ref{dpfg}) is poor. 

If either $R\gtrsim\lambda_L$ or $L\gtrsim\lambda_L$, then the interference of radiation of all particles is constructive inside a cone with an opening angle $\delta\theta\lesssim1$ and destructive elsewhere, implying that the radiation is coherent only inside this narrow cone. The value of $\delta\theta$ can be identified by demanding the coherence factor Eq.~(\ref{alphag})   essentially non vanishing for $\omega\sim\omega_L$, that is $\delta\theta\sim\lambda_L/R$ for $R^2/L\lambda_L\gg 1$ and  $\delta\theta\sim\sqrt{\lambda_L/L}$ for $R^2/L\lambda_L\ll 1$. Since $\left<\Delta p_\parallel\right>\sim\delta\theta^4$ for $\delta\theta\lesssim 1$, see Eq.~(\ref{dP-}), this way we deduce Eqs.~(\ref{ggb}) and (\ref{ggc}) up to constant numerical  factors directly from Eq.~(\ref{gga}).

\begin{figure}
    \centering    \includegraphics[width=0.95\linewidth]{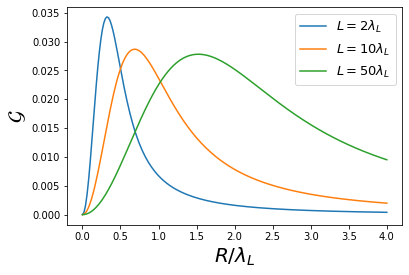}
    \caption{Geometric factor $\mathcal{G}$, Eq.~(\ref{ggint}) vs bunch width $R$ for three values of the bunch thickness $L$.}    \label{fig_Gg}
\end{figure}

From Eqs.~(\ref{ggb}) and (\ref{ggc}) it follows that the geometric factor $\mathcal{G}$ of a large bunch (either $R\gtrsim\lambda_L$ or $L\gtrsim\lambda_L$) has a maximum at a certain value of the parameter $\rho\sim R^2/L\lambda_L$, see Fig.~\ref{fig_Gg}. Let us estimate it for a thick bunch with $L\gtrsim\lambda_L$. First, observe that for $\rho\gg1$ the geometric factor is suppressed as $\mathcal{G}\propto\rho^{-1}\ll1$. If $\rho\lesssim 1$, then the second term in the exponential in Eq.~(\ref{ggint})  can be neglected, as only $x\lesssim 1$ effectively contribute to the integral due to the first term therein. Therefore,
\begin{equation}\label{G0}
\mathcal{G}\approx\frac{\rho}{4\pi}\left[1-\sqrt{\frac{\pi}{2}}\rho e^{\frac{\rho^2}{2}}\mathrm{erfc}\left(\frac{\rho}{\sqrt{2}}\right)\right],
\end{equation}
where $\mathrm{erfc}(z)=1-\mathrm{erf}(z)$ is the complementary error function. The numerical maximization of Eq.~(\ref{G0}) provides
\begin{equation}\label{Gm}
    \mathcal{G}^{m}\approx0.03\quad \textrm{at}\quad \frac{R^2}{L\lambda_L}\approx 0.05.
\end{equation}
Direct evaluation of the integral Eq.~(\ref{ggint}) confirms the estimate of its maximum and the optimal shape $R\sim\sqrt{L\lambda_L}$, see Fig.~\ref{fig_Gg}.

The geometric factor Eq.~(\ref{ggint}) can be expressed in a closed form as 
\begin{equation}\label{Gg}
    \mathcal{G}=\frac{\sqrt{A}B\left(e^{A^2}-e^{B^2}+\sqrt{\pi}B\left[\mathrm{erfi}(B)-\mathrm{erfi}(A)\right]\right)}{2\sqrt{2}\pi e^{  B^2}\sqrt{B-A}},
\end{equation}
where $A=\pi L^2/\left[2\lambda_L\sqrt{2(16R^2-L^2)}\right]$ and $B=4\sqrt{2}\pi R^2/(\lambda_L\sqrt{16R^2-L^2})$ and $\mathrm{erfi}(z)$ is the imaginary error function. 

Now let us turn to the backward contribution $\left<\Delta p^b_\parallel\right>$. Substituting Eqs.~(\ref{C}), (\ref{alphag}) and (\ref{dE1b}) into Eq.~(\ref{dP-}) and replacing $\cos\theta\to-1$, $\sin(\theta/2)\to1$, we obtain
\begin{equation}\label{dpbg0}
\begin{split}
    &\left<\Delta p^b_\parallel\right>\approx mc\frac{\sqrt{\pi}}{16}\frac{n}{n_c}\frac{\omega_L^4 R^2LT}{c^3} \mathcal{I} ,\\
    &\mathcal{I}=\int e^{-\frac{\omega^2}{2c^2}(\frac{L^2}{4}+R^2\sin^2\theta)}\frac{1+\sin\eta}{\nu}\frac{\omega d\omega}{\omega_L^2}\sin\theta d\theta,
\end{split}
\end{equation}
where $\eta$ and $\nu$ are defined in Eq.~(\ref{eta}). Here the integration over frequency effectively starts from the vicinity of the position $\omega_1$ of the first peak in the single particle radiation spectrum as defined in Eq.~(\ref{om1}) [see Fig.~\ref{fig_backspect} and its discussion].
On the opposite side, integration formally extends to infinity, but Eq.~\eqref{dE} in use is valid only for $\omega<\omega_*$. Nevertheless, under quite a weak condition $\omega_* L/c\gg1$ the integrand falls out exponentially already within the applicability range, so that we can still use Eq.~\eqref{dE} integrating up to infinity. 

As repeatedly emphasized, for our analytical approach we assume that the radiation is emitted in a narrow angle, i.e. a short angular range contributes to the integral. As suggested by Eq.~\eqref{dpbg0}, this is indeed the case, if  $\omega_1 R/c\gg1$. Then, expanding $\sin\theta\approx\pi-\theta$, integrating over $\theta$ and using  $\omega\approx\omega_1 e^{\nu^2/2}$ (which is valid for $\omega\ll\omega_*$), we get
\begin{equation}\label{I1}    \mathcal{I}\approx\frac{c^2}{\omega_L^2R^2}\int\limits_0^\infty e^{-\frac{\omega_1^2 L^2 e^{\nu^2}}{8c^2}}(1+\sin\eta) d\nu.
\end{equation}
The resulting integral in Eq.~\eqref{I1} can be approximated analytically assuming the bunch is thick enough. In this case the effectively contributing $\nu$ are small $\nu\lesssim c/(\omega_1 L)$, so that we can expand $e^{\nu^2}\approx 1+\nu^2$ and $\eta\approx \omega_L T\nu^3/3$. Finally, expanding also $\sin{\eta}$ and assuming $\omega_1L/c\gtrsim1$ and $\omega_1L/c\gtrsim (\omega_L T)^{1/3}$,  we obtain 
\begin{equation}\label{dpbg}
\left<\Delta p^b_\parallel\right>\approx mc\frac{\pi\omega_L T}{32\sqrt{2}}\frac{a_0^2}{\gamma_0^2}\frac{n}{n_c}e^{-\frac{\pi^2L^2}{2\lambda_1^2}},
\end{equation}
where $\lambda_1\approx\frac{\lambda_L(1+a_0^2)}{4\gamma_0^2}$ is the wavelength of the fundamental harmonic of the backward radiation.

Let us represent the total average momentum loss due to coherently enhanced RF in the form 
\begin{equation}\label{dpt}
    \left<\Delta p_\parallel\right>\approx\left<\Delta p^f_\parallel\right>(1+\upsilon), \quad \upsilon=\frac{\left<\Delta p^b_\parallel\right>}{\left<\Delta p^f_\parallel\right>}=\frac{e^{-\frac{\pi^2L^2}{2\lambda_1^2}}}{4\pi\mathcal{G}},
\end{equation}
where $\left<\Delta p^f_\parallel\right>$ is defined in Eq.~(\ref{dpfg}). For a thick bunch, $L\gg\lambda_1$, the backward radiation is suppressed, $\upsilon\ll1$, and only forward radiation contributes to RF.

Being proportional to the particle density, the effect is essential for a dense bunch with $n\sim n_c$ [for an optical laser $n_c\sim10^{21}$cm$^{-3}$] or few orders of magnitude lower, see the companion paper Ref.~\cite{prl_arxiv}. Quasi--monoenergetic bunches with $n\sim10^{18}-10^{19}$ cm$^{-3}$ have been already obtained experimentally \cite{salehi_prx2021,chang_prappl2023,storey_prstab2024} and some  ways to generate bunches with $n\sim10^{21}$cm$^{-3}$ and even higher have been discussed in e.g. Refs.~\cite{yakimenko_prl2019,facet2,shi_prl2021,shi_ppcf2021,shi_hplse2022,blackman_commphys2022}.

Three dimensional particle-in-cell simulations reported in the companion paper Ref.~\cite{prl_arxiv} demonstrate high accuracy of the approximation Eq.~(\ref{dpt}) for $ \left<\Delta p_\parallel\right>\ll p_0$. If this inequality is violated, this means that RF is so strong that significantly perturbs the particle motion. In such a case the expressions for the single particle radiation spectra Eqs.~(\ref{dE1f}), (\ref{dE1b}) obtained by neglecting the effect of RF on the particle trajectories are no longer valid, at least quantitatively.

\section{Discussion}\label{sec_disc}

In order to compare the effect of  coherently enhanced RF Eq.~(\ref{dpt}) with the known single-particle result Eq.~(\ref{deltapll}), let us write their ratio as 
\begin{equation}\label{ratio2}
    \frac{\left<\Delta p_\parallel\right>}{\Delta p^{LL}_\parallel}\approx \frac{\pi^{3/2}}{16}\frac{n}{n_c}\frac{(1+\upsilon)\mathcal{G}}{\mu\gamma_0^4}.
\end{equation}
The denominator contains $\mu$, which for an optical/infrared laser is small, $\mu\sim10^{-8}-10^{-9}$. As a consequence, RF can be coherently enhanced at low frequencies by orders of magnitude \cite{gelfer_prr2024} by using mildly relativistic particle bunches. This is clearly seen by comparing the single-particle radiation spectrum in Fig.~\ref{fig_rad}~(b), which corresponds to the per-particle incoherent contribution, to the actual total radiation spectrum in Fig.~\ref{fig_cohspect}.  

On the other hand, due to the presence of the Lorentz factor in the denominator of Eq.~\eqref{ratio2}, for ultrarelativistic bunches the incoherent contribution to RF always overcomes the coherent one. This is explained by that for $a_0\gtrsim1$ and $\gamma\gg1$ the incoherent single-particle spectrum is much broader, extending up to  frequencies $\omega\lesssim a_0 (a_0^2+\gamma_0^2)\omega_L$, see Fig.~\ref{fig_rad}~(b) and Refs.~\cite{jackson_book,ritus1985,esarey_pre1993}. Further  comparison of the properties of  ordinary and coherently enhanced RF can be found in the companion paper \cite{prl_arxiv}.

\begin{figure}
    \centering    \includegraphics[width=0.95\linewidth]{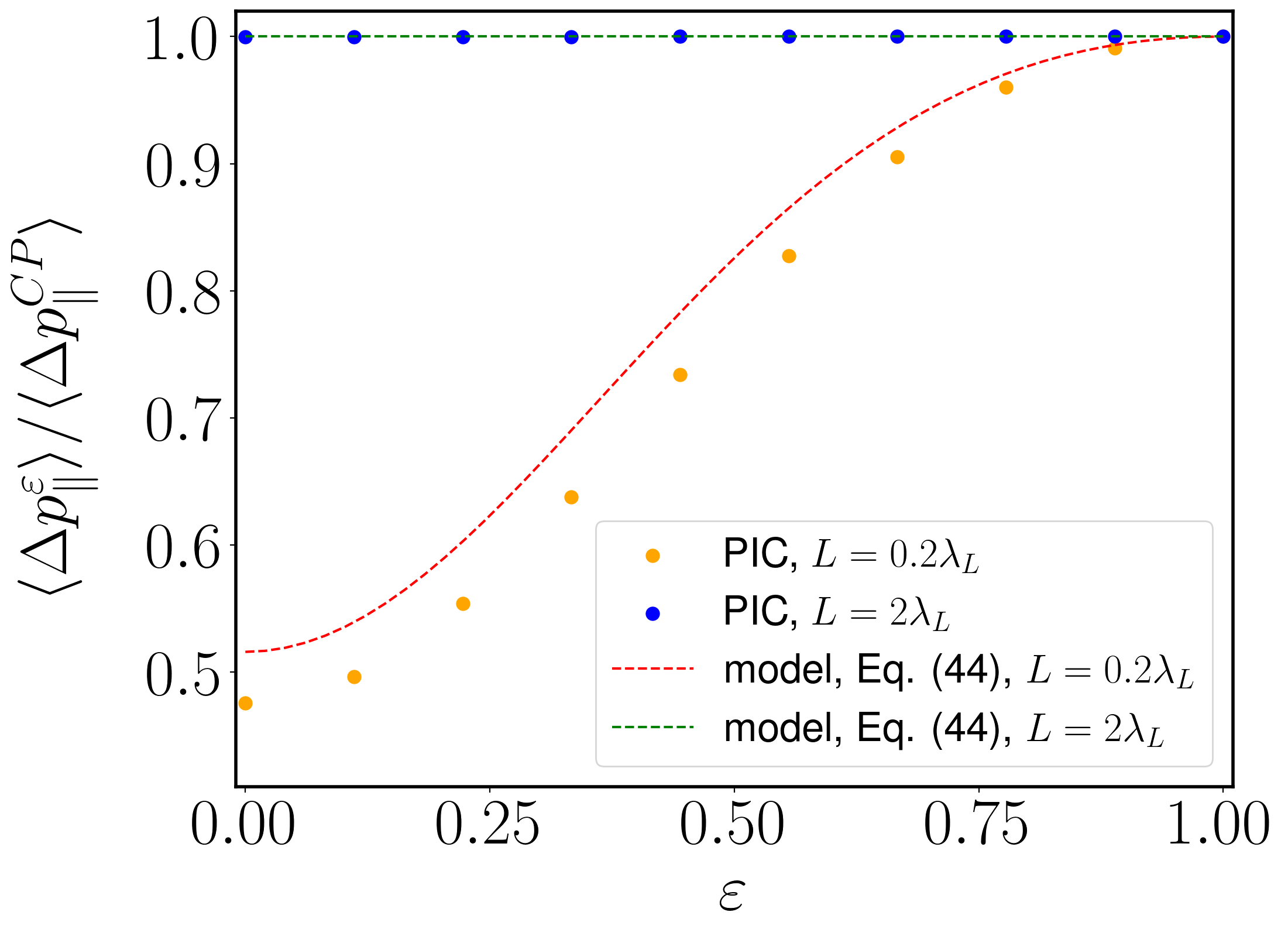}
    \caption{Dependence of the momentum loss on laser polarization. Parameters: $a_0=\gamma_0=5,\  T\approx 17$ fs, $R=2\lambda_L, n=0.1 n_c$.}    \label{fig_elliptic}
\end{figure}

Next let us study the sensitivity of our results with respect to the  simplifications in use, such as the assumed circular  polarization (CP) Eq.~(\ref{A}) of the laser pulse and the specific (Gaussian) density profile of the particle bunch.

For an arbitrary elliptical polarization
\begin{equation}\label{Ael}
\mathbf{A}=\frac{mca_0}{e}\sqrt{\frac{2}{1+\varepsilon^2}}e^{-\phi^2/(\omega_LT)^2}\{0,\cos\phi,\varepsilon\sin\phi\},
\end{equation}
the single particle spectrum in the forward direction (\ref{dE1}) coincides with the CP result Eq.~(\ref{dE1f}), i.e. is independent of polarization $\varepsilon$. As for the backward spectrum,  one can demonstrate \cite{malakhov_prep} that for $a_0\gg1$ in the vicinity of the first harmonic $\omega\sim\omega_1$ that mostly contributes to the coherent radiation,
\begin{equation}
\begin{split}
     &\left.\frac{d\mathcal{E}^{(1)}}{d\omega d\Omega}\right|_{\theta=\pi}^\varepsilon\approx\mathcal{R}_\varepsilon\left.\frac{d\mathcal{E}^{(1)}}{d\omega d\Omega}\right|_{\theta=\pi}^{\varepsilon=1},\ \mathcal{R}_\varepsilon=1-\frac{1}{2}\left(\frac{1-\varepsilon^2}{1+\varepsilon^2}\right)^2,
\end{split}
\end{equation}
where 
$\left.d\mathcal{E}^{(1)}/d\omega d\Omega\right|_{\theta=\pi}^{\varepsilon=1}$ is the single particle backward spectrum for CP, see Eq.~(\ref{dE1b}). 

Thus, since the coherence factor $\alpha$ is obviously independent of $\varepsilon$, the effect of polarization on the coherently enhanced RF is given by
\begin{equation}\label{dpeps}
    \frac{\left<\Delta p_\parallel^\varepsilon\right>}{\left<\Delta p_\parallel^{CP}\right>}\approx \frac{1+\mathcal{R}_\varepsilon\upsilon}{1+\upsilon},
\end{equation}
where $\left<\Delta p_\parallel^\varepsilon\right>$ is the momentum loss in an elliptically polarized pulse, $\left<\Delta p_\parallel^{CP}\right>$ is the CP result Eq.~(\ref{dpt}) and the ratio between the backward and forward contributions $\upsilon$ in the CP case was defined in Eq.~(\ref{dpt}). 

According to Eq.~(\ref{dpeps}), for a thick bunch such that the backward contribution is negligible,  $\upsilon\ll1$ and RF is independent of polarization. In the opposite case $\upsilon\gg1$ the effect for the linear polarization $\varepsilon=0$ is approximately twice weaker than for CP. This is in agreement with the results of PIC simulations shown in Fig.~\ref{fig_elliptic}, see the companion paper \cite{prl_arxiv} for the description of the numerical approach. 

To check the dependence on the initial density profile of the bunch, we consider an alternative bunch model of a uniform cylinder of length $L$ and radius $R$. The analytical calculation follows the same lines as for the Gaussian density distribution Eq.~(\ref{alphag}) in Secs.~\ref{sec_rad} and \ref{sec_mom}, see Appendix~\ref{app_rect} for details. We found the results for these two density distributions  quite similar, meaning that the effect is robust with respect to the particular shape of the bunch, which is hard to control in experiments anyway. 

\begin{figure}
    \centering    \includegraphics[width=0.95\linewidth]{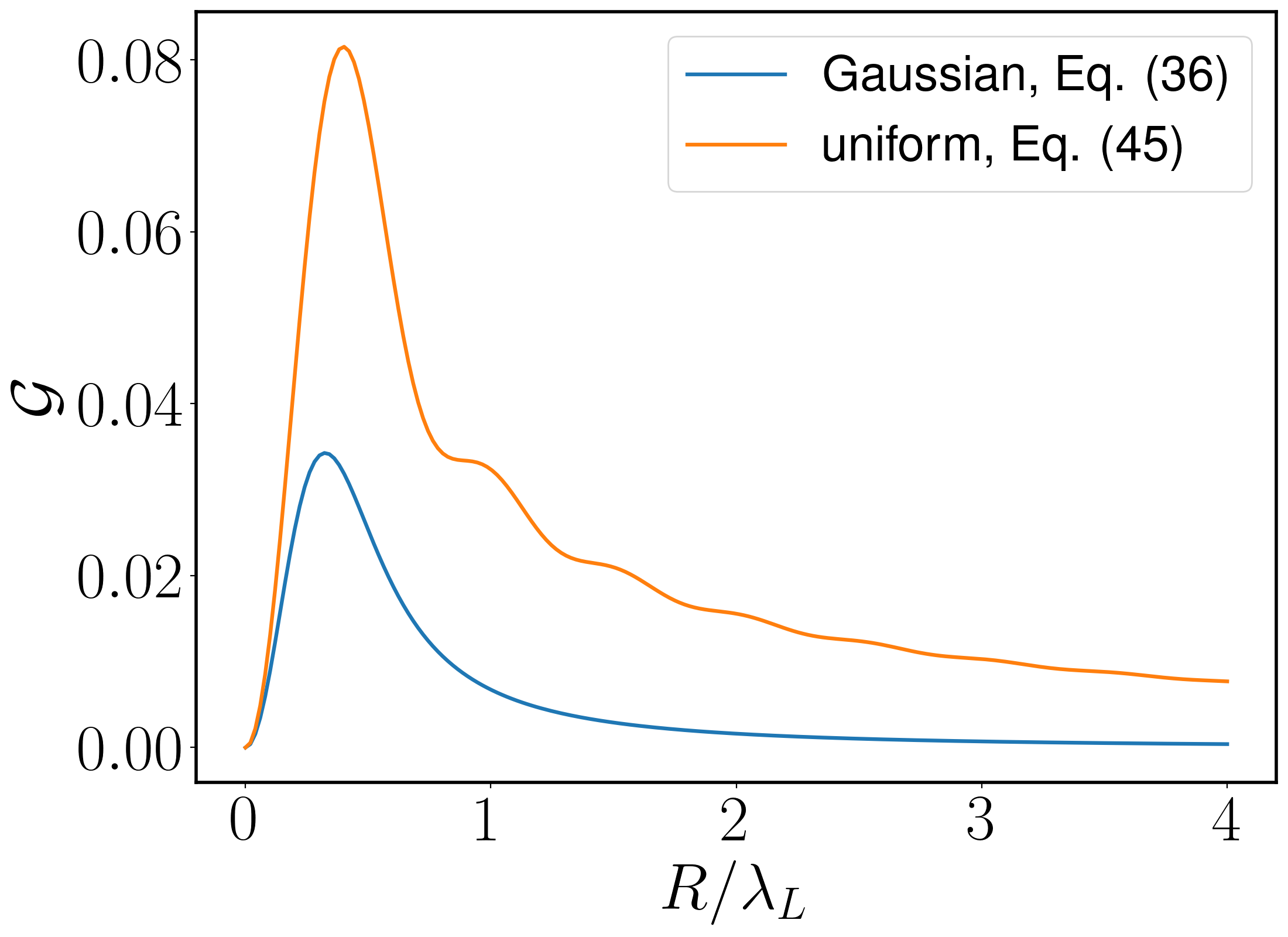}
    \caption{Geometric factors for Gaussian and uniform distributions of the bunch density.}   \label{fig_Gguni}
\end{figure}

In particular, the forward contribution to RF can be represented by the same Eq.~(\ref{dpfg}) as in the Gaussian case, but with a different geometric factor 
\begin{equation}\label{guint}
\mathcal{G}_u=\pi^{-3/2}\int\limits_{0}^{\frac{\pi L}{2\lambda_L}}\frac{J_1^2\left(\sqrt{2\rho x\left(1-\frac{\lambda_L}{\pi L}x\right)}\right)\sin^2x dx}{x^2\left(1-\frac{\lambda_L}{\pi L} x\right)},
\end{equation}
where $\rho$ and $x$ are the same as defined below Eq.~(\ref{ggint}). It is about twice larger than the Gaussian one but otherwise behaves similarly, see Fig.~\ref{fig_Gguni}, in particular reaching its maximum at $R^2/L\lambda_L\approx 0.05$, as in Eq.~(\ref{Gm}). 

\section{Summary}\label{sec_sum}

We have studied radiation losses of a dense particle bunch colliding head-on with a laser pulse. We developed an analytical model, which describes the average momentum losses of particles in the bunch in dependence on various laser and bunch parameters, such as laser pulse intensity, duration and polarization, and bunch size, density and initial energy. The model unfolds the conditions for a strong coherent enhancement of radiation in the forward and backward directions at low frequencies, thus also for enhancement of radiation friction. It is shown that the results are marginally sensitive to laser polarization and density distribution of the particle bunch. Numerical simulations supporting our analytical findings (including the effect of the neglected here momentum spread of the particles in the bunch) and further discussion of physical origins of the coherently enhanced radiation friction, its distinction from the ordinary noncoherent radiation friction, and the prospects for experimental observation can be found in the companion paper Ref.~\cite{prl_arxiv}.

\section*{Acknowledgments}
The authors are grateful to Antonino Di Piazza, Matteo Tamburini, Mickael Grech, Caterina Riconda, Sergey Rykovanov, Igor Kostyukov, Vladimir Tikhonchuk, Martin Stack Formanek and Peter Valenta for valuable discussions. E.G.G., J.C., O.K. and S.W. were supported by NSF--GACR project 24-14395L. A.M.F. was supported by the Russian Science Foundation (Grant No. 25-12-00336). The numerical simulations were performed using the code SMILEI and the resources of the ELI ERIC SUNRISE cluster. 

\appendix
\section{Radiation and radiation friction of a uniformly distributed particle bunch}\label{app_rect}
To test the robustness of our results, consider a cylindrical particle bunch of length $L$, radius $R$ with a uniform density distribution $n(x,\mathbf{r}_\bot)=n_0\theta(x)\theta(L-x)\theta(R-r_\bot)$, where $\theta(x)$ is the Heaviside step function. Then for the coherence factor  $\alpha=|\left<e^{i\Phi}\right>|^2$ we obtain~\footnote{Note that there is a typo in the expression for $\alpha$ for this case in Ref.~\cite{gelfer_prr2024}, where $L$ should be replaced with $L/2$.}
\begin{equation}\label{alphau}
\alpha=\frac{16 c^4 J_1^2\left(\frac{\omega R}{c}\sin\theta\right)\sin^2\left(\frac{\omega L}{2c}\sin^2\frac{\theta}{2}\right)}{ R^2L^2\omega^4\sin^2\theta\sin^4\frac{\theta}{2}},
\end{equation}
where $J_1(x)$ is the Bessel function and the initial phases $\Phi_j$ are  defined as in Eq.~(\ref{phij}).

Let us calculate the momentum transfer Ref.~(\ref{dP-}) for a uniform bunch, again considering the forward and backward contributions separately. While evaluating the backward contribution $\left<\Delta p_\parallel^{b}\right>$, we assume that the bunch size is larger than the wavelength of the lowest backward scattered harmonic and that the pulse is longer than at least a few laser periods,
\begin{equation}\label{ubcond}
\omega_1 R/c\gtrsim 1,\quad  \omega_1 L/c \gtrsim 1, \quad \omega_L T\gg1.
\end{equation}
Note that this does not necessarily imply that the bunch is larger than the laser wavelength, as in our setup $\omega_1>\omega_L$ for $\gamma_0>a_0$.

Under the conditions (\ref{ubcond}) the backward radiation sharply peaks around $\theta=\pi$ \cite{gelfer_prr2024}, and, as before, we substitute into Eq.~\eqref{dP-} $\sin(\theta/2)\to1$, $\cos\theta\to -1$ and use $d\mathcal{E}^{(1)}/d\omega d\Omega$ from Eq.~(\ref{dE1b}). Furthermore, in the integral over frequency we replace the lower limit with $\omega_1$ where starts the single particle spectrum, and the upper limit with infinity due to a fast fall out of $\alpha\propto\omega^{-5}$ at high frequencies, see Eq.~(\ref{alphau}). 

With all the above adjustments we obtain
\begin{equation}\label{dpmb1}
\left<\Delta p_\parallel^{b}\right>\approx \frac{4Ne^2c^3T}{R^2L^2}\int\limits_{\omega_1}^\infty\frac{1+\sin\eta}{\omega^3\nu}\sin^2\frac{\omega L}{2c}d\omega,
\end{equation}
in particular by taking into account that
\begin{equation}\label{Jint}
\int\limits_{\pi/2}^\pi\frac{J_1^2\left(\frac{\omega R}{c}\sin\theta\right)}{\sin\theta}d\theta\approx\frac{1}{2},
\end{equation}
for $\omega R/c\gtrsim 1$, when the angular range $\pi-c/\omega R\lesssim\theta<\pi$ effectively contributing to the integral in Eq.~(\ref{Jint}) is narrow \footnote{For example, for $\omega>\omega_L$ and $R>\lambda_L/2$ the accuracy of Eq.~(\ref{Jint}) is better than 6\%.}. 

\begin{figure}[t]
\includegraphics[width=0.45\textwidth]{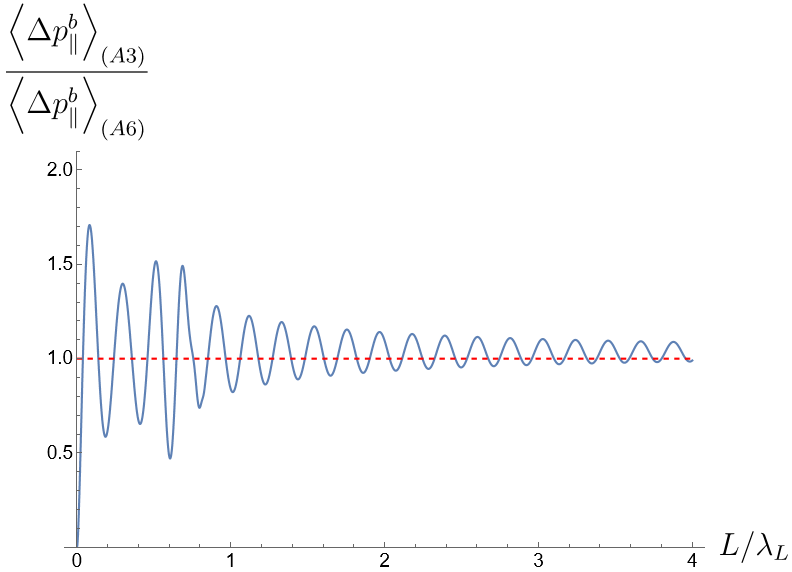}
\caption{Ratio of the backward contribution to the momentum loss evaluated numerically via Eq.~(\ref{dpmb1}) and analytically via Eq.~(\ref{dp1b}) for $\omega_L T=5\pi$.}\label{fig_dpmb}
\end{figure}

To evaluate the remaining integral over frequency, we change the integration variable to $\nu$ and observe that, since the largest contribution comes from the lower integration limit, we can approximate $\nu\approx\sqrt{2\ln(\omega/\xi^2\omega_L)}$ for $a_0\gg 1$, thus obtaining
\begin{equation}
\left<\Delta p_\parallel^{b}\right>\approx \frac{4Ne^2c^3T}{\omega_L^2R^2L^2\xi^4}\int\limits_{\sqrt{2\delta}}^\infty(1+\sin\eta)e^{-\nu^2}\sin^2\Lambda d\nu,
\end{equation}
where $\Lambda=\xi^2\omega_L Le^{\nu^2/2}/2c$ and $\delta$ is defined in Eq.~\eqref{om1}. 

In the remaining integral, we assume that $\xi^2\omega_L L/c\gg 1$ and average the integrand over the oscillations of $\sin^2\Lambda$ by replacing the latter with $1/2$. To evaluate the term with $\sin\eta$ we observe that the main contribution to the integral comes from  frequencies $\omega\ll\omega_*$ and Taylor-expand $\eta\approx \omega_L T\nu^3/3$. Furthermore, assuming $\omega_L T\gg 1$, we see that $\sin(\omega_L T\nu^3/3)$ undergoes rapid oscillations suppressing the integral unless $\nu\ll 1$, hence in this term approximate $e^{-\nu^2}\approx 1$. The other term is evaluated directly.
Thus we arrive at
\begin{equation}\label{dp1b}
\begin{split}
&\left<\Delta p_\parallel^b\right>\approx\frac{mc a_0^4}{128\sqrt{\pi}\gamma_0^4}\frac{n}{n_c}\frac{\lambda_L}{L}\mathcal{T}(\omega_L T),\\
&\mathcal{T}(t)=t\left\{\mathrm{erfc}\left[\left(\frac{3\pi}{2t}\right)^{1/3}\right]+\beta t^{-1/3}\right\},
\end{split}
\end{equation}
where $\beta=(2/3^{2/3}\sqrt{\pi})\int_{\pi/2}^\infty\sin\eta\, \eta^{-2/3} d\eta\approx 0.097$ and we took into account that $N=\pi nR^2L$. Note that for $\omega_L T\to \infty$ we have $\mathcal{T}(\omega_L T)\to\omega_L T$. However, in numerical simulations below we take $\omega_L T=5\pi$ (as corresponding to $T\approx 17 \mathrm{fs}$), for which $\mathcal{T}(\omega_L T)\approx 0.38$. For this reason in the sequel we avoid further simplifications of the function $\mathcal{T}(\omega_L T)$. 

To check the validity of the approximations resulted in analytical result Eq.~(\ref{dp1b}) we benchmark it against the numerical evaluation of Eq.~(\ref{dpmb1}). Figure~\ref{fig_dpmb} demonstrates that Eq.~(\ref{dp1b}) is fairly accurate for $L\gtrsim\lambda_L$ and $\omega_L T=5\pi$ as used in numerical simulations below. The oscillations in Fig.~\ref{fig_dpmb} clearly appear due to the actual oscillations of the factor $\sin^2(\omega L/2c)$ in Eq.~(\ref{dpmb1}) and a small discrepancy in the mean values arises due to our approximation $e^{-\nu^2}\approx 1$ in the evaluation of the term that involves $\sin\eta$. 

\begin{figure}[t]
\includegraphics[width=0.45\textwidth]{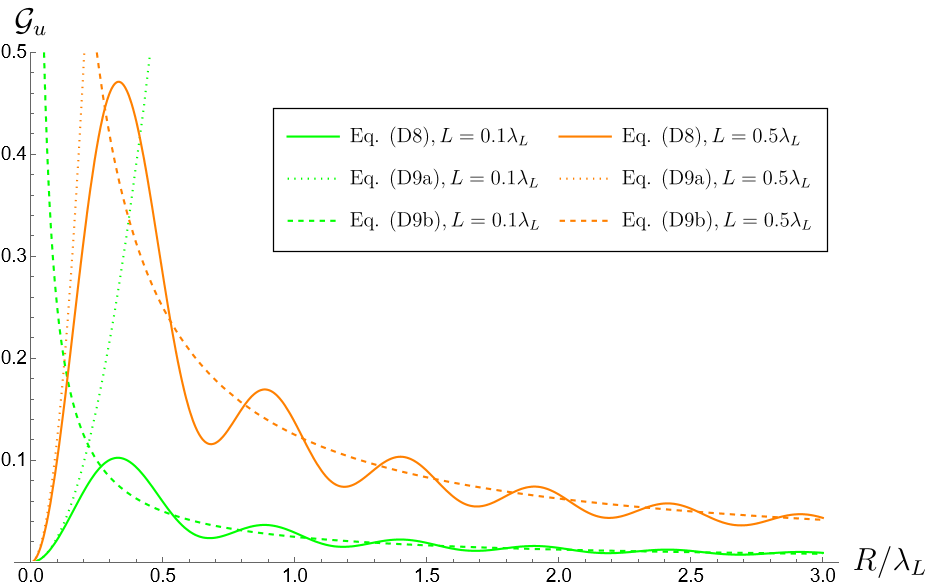}
\includegraphics[width=0.45\textwidth]{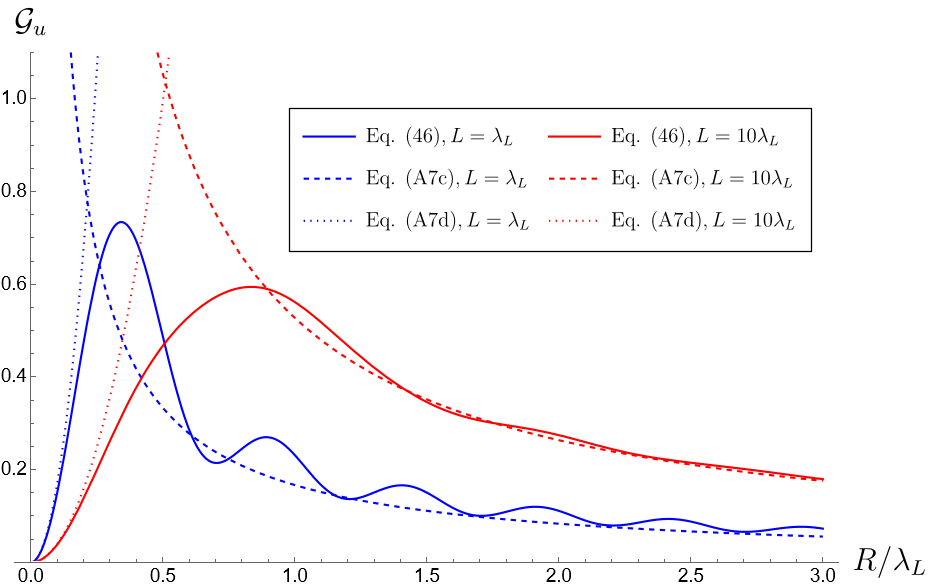}
\caption{Comparison of the analytical approximations (\ref{grl}) for the geometric factor $\mathcal{G}$ [dotted and dashed lines] versus direct numerical evaluation of Eq.~(\ref{guint}) [solid lines] for several bunch lengths $L$. Green, orange, blue and red colors correspond to $L/\lambda_L=0.1$, $0.5$, $1$ and $10$, respectively.}\label{fig_gr}
\end{figure}

For the forward contribution $\left<\Delta p_\parallel^f\right>$ to the momentum loss, we obtain an expression similar to the Gaussian case (\ref{dpfg}), but with a different geometric factor Eq.~(\ref{guint}). Unlike the Gaussian case, here the integral representing the geometric factor $\mathcal{G}_u$ cannot be evaluated exactly. However, it can still be approximated analytically under further assumptions on the bunch sizes $R$ and $L$:
\begin{subequations}\label{grl}
\begin{align}
&\mathcal{G}_u(L\ll\lambda_L, R\ll\lambda_L)=\frac{\pi^{3/2}}{2}\frac{R^2L}{\lambda_L^3}\label{grla},\\
&\mathcal{G}_u(L\ll\lambda_L, R\gg\lambda_L)=\frac{1}{2\pi^{5/2}}\frac{L}{R}\label{grlb},\\
&\mathcal{G}_u\left(L\gtrsim\lambda_L, \frac{R^2}{L\lambda_L}\gg 1\right)=\frac{1}{3\pi^{5/2}}\frac{\sqrt{L\lambda_L}}{R},\label{grlw}\\
&\mathcal{G}_u\left(L\gtrsim\lambda_L, \frac{R^2}{L\lambda_L}\ll 1\right)=\frac{2}{\sqrt{\pi}}\frac{R^2}{\lambda_L L}\left(C+\ln\frac{\pi L}{\lambda_L}\right),\label{grln}
\end{align}
\end{subequations}
where $C$ is the Euler constant.

For a thin bunch $L\ll \lambda_L$ we pass from Eq.~(\ref{guint}) to Eqs.~(\ref{grla}), (\ref{grlb}) by approximating $\sin^2x/x^2\approx 1$ and expanding the Bessel function for either small (if $R\ll \lambda_L$) or large (if $R\gg\lambda_L$) argument, respectively. Again, similar to the Gaussian case, if both sizes of the bunch are much smaller than the laser wavelength, then
$\left<\Delta p_\parallel^f\right>$ is proportional to $N$ and independent of the shape of the bunch, simply meaning that in this case all the particles in the bunch emit coherently at $\omega\sim\omega_L$.

In the opposite case $L\gtrsim \lambda_L$, the major contribution to the integral in Eq.~(\ref{guint}) comes from $x\lesssim 1$ due to the factor $\sin^2 x/x^2$. Physically, this restricts the width of the radiation angular distribution \cite{gelfer_prr2024}. Hence we Taylor expand in powers of $\lambda_L/\pi L<1$ and also expand the Bessel function for either large or small argument with respect to $\rho$, thus arriving at Eq.~(\ref{grlw}) or Eq.~(\ref{grln}). Figure~\ref{fig_gr} demonstrates the accuracy of the resulting analytical approximations (\ref{grl}). Note that for $L\ll\lambda_L$  the geometric factor $\mathcal{G}_u$ attains its maximum at $R\sim\lambda_L/3$   independently of the value of $L$ (Fig.~\ref{fig_gr}, top), while for $L\gtrsim\lambda_L$ (Fig.~\ref{fig_gr}, bottom) the position of the maximum increases with $L$ in agreement with Eq.~(\ref{Gm}). 

\begin{figure}[t]
\includegraphics[width=0.45\textwidth]{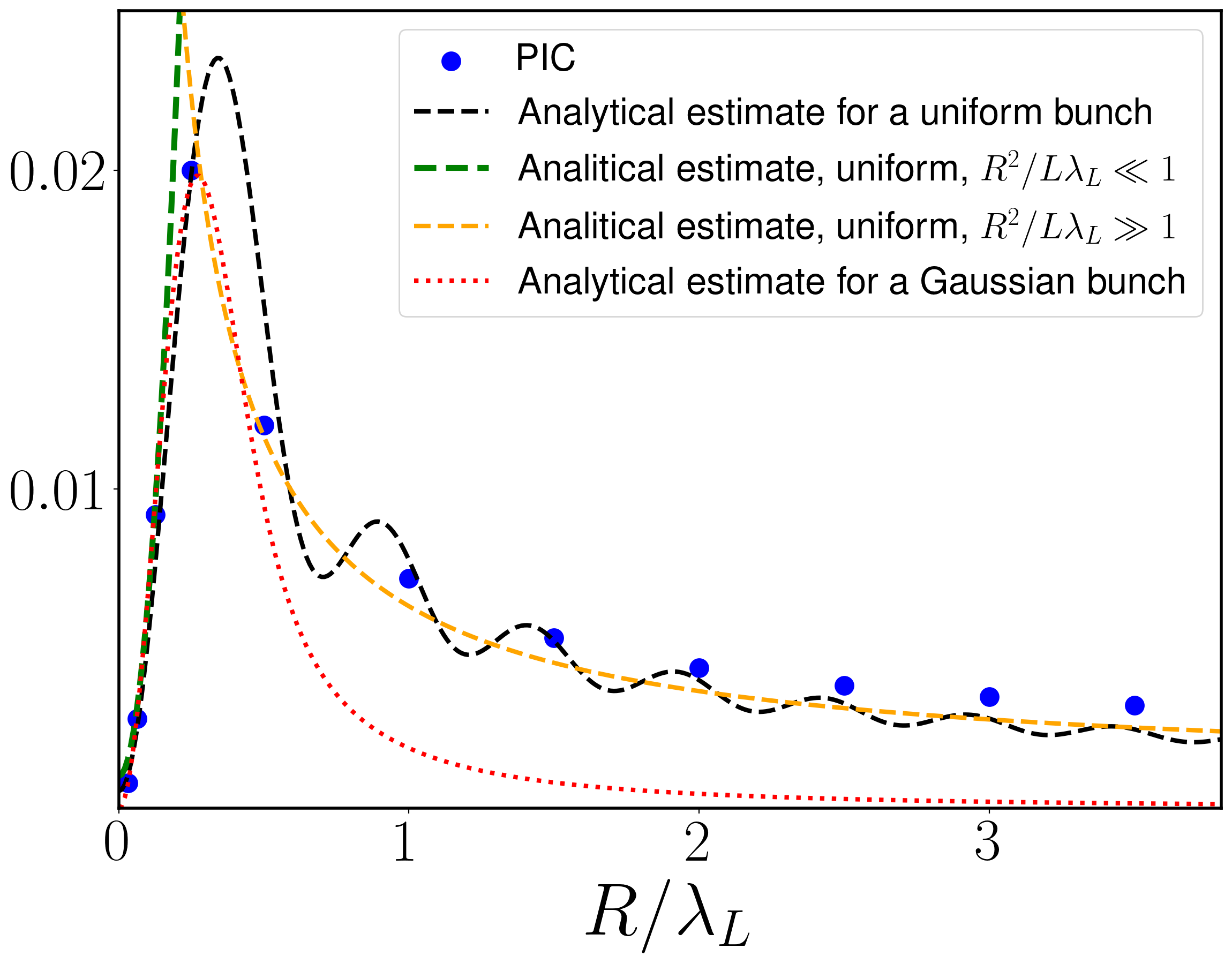}
\caption{Normalized particle deceleration $\left<\Delta p_\parallel\right>/p_0$ vs bunch width $R$ for a cylindrical bunch of uniform density distribution. Blue dots are extracted from 3D PIC simulations, dashed curves correspond to the sum Eq.~(\ref{dpfg})+Eq.~(\ref{dp1b}) with  $\mathcal{G}$ from Eqs.~(\ref{guint})/(\ref{grlw})/(\ref{grln}) -- black/orange/green; red dotted curve corresponds to the Gaussian estimate (\ref{dpfg}) with $\mathcal{G}$ from Eq.~(\ref{Gg}) multiplied by a scale factor $2$. Other parameters: $a_0=5$, $\omega_L T=5\pi$, $n=0.1 n_c$, $L=\lambda_L$. }\label{fig_results_uni}
\end{figure}

Figure~\ref{fig_results_uni} illustrates the dependence of the coherent RF on a transverse size $R$ of a bunch with a uniform initial density as extracted from 3D PIC simulations (the numerical approach is described in the companion paper \cite{prl_arxiv}). In accordance with Eq.~(\ref{grl}) one can clearly see a crossover between the small and large $R$ asymptotics where RF is maximized. Its position at the intersection of the green and purple curves matches the condition (\ref{Gm}). As we have already noted, the numerical value $0.05$ of the constant in Eq.~(\ref{Gm}) turns out to be close for the uniformly and Gaussian distributed bunches. It is worth stressing that whereas the backward contribution can be actually neglected for small $R$, taking it into account for large $R$ considerably improves the accuracy of the model. 

Finally, let us compare the results for the uniform and Gaussian density distributions. Whereas in the Gaussian case for $L>\lambda_1$, where $\lambda_1\approx\hat{\lambda}$ is the wavelength of the first harmonic in the backward radiation spectrum, the backward contribution (\ref{dpbg}) falls out exponentially, for a uniform distribution it falls out as $1/L$, see Eq.~(\ref{dp1b}). In addition, in contrast to the former case, its dependence on $L$  actually becomes oscillatory. The origin of these oscillations can be understood as follows. Backward radiation from particles in a slice of thickness $\lambda_1/2$ interferes constructively, but destructively from such adjacent slices. Therefore, if the thickness of the whole bunch $L=m\lambda_1$, where $m$ is an integer, then the backward radiation vanishes, while it is maximal for $L=(m+1/2)\lambda_1$. These oscillations correspond to the factors $\sin^2(\omega L\sin^2(\theta/2)/2c)$ in Eq.~(\ref{alphau}) and $\sin^2(\omega L/2c)$ in Eq.~(\ref{dpmb1}); for simplicity we averaged them out in our final expression (\ref{dp1b}). 

For a thick ($L\gtrsim\lambda_L$) bunch and $R^2/L\lambda_L\lesssim1$ the forward contribution dominates over the backward one for both uniform and Gaussian density distributions. In this regime the scaling of RF with laser and bunch parameters for both distributions is in fact quite similar, cf. Eqs.~(\ref{ggc}) and (\ref{grln}). Figure~\ref{fig_results_uni} demonstrates that for $R\lesssim\sqrt{L\lambda_L}$ the scaling for a Gaussian bunch almost perfectly match the simulation results for a uniform bunch up to a constant scaling factor. As for  $R\gg\sqrt{L\lambda_L}$, the forward contribution $\left<\Delta p_\parallel^f\right>$ decreases as $R^{-2}$ in the Gaussian case and as $R^{-1}$ in the uniform case, so that at some point the backward contribution can become dominant in both cases.

\bibliography{lit_arxiv}

@article{dipiazza_rmp2012,
    author = {Di Piazza, A and M{\"u}ller, C and Hatsagortsyan, KZ and Keitel, Ch H},
    doi = {10.1103/RevModPhys.84.1177},
    journal = {Rev. Mod. Phys.},
    number = {3},
    pages = {1177},
    publisher = {APS},
    title = {Extremely high-intensity laser interactions with fundamental quantum systems},
    volume = {84},
    year = {2012}
}

@book{landau2,
    author = {Landau, Lev Davidovich and Lifshitz, I M},
    publisher = {Course of Theoretical Physics Series, Pergamon Press, London},
    title = {Theoretical Physics: The Classical Theory of Fields},
    year={1988},
    volume = {2}
}

@article{ritus1985,
  	title={Quantum effects of the interaction of elementary particles with an intense electromagnetic field},
	author={Ritus, V I},
	journal={J. Russ. Laser Res.},
	volume={6},
	number={5},
	pages={497--617},
	year={1985},
	publisher={Springer},
	doi={https://doi.org/10.1007/BF01120220}
	}

@article{schwinger1951,
	title={On gauge invariance and vacuum polarization},
	author={Schwinger, Julian},
	journal={Phys. Rev.},
	volume={82},
	number={5},
	pages={664},
	year={1951},
	publisher={APS},
	doi={10.1103/PhysRev.82.664}
}

@article{cole_prx2018,
  title = {Experimental Evidence of Radiation Reaction in the Collision of a High-Intensity Laser Pulse with a Laser-Wakefield Accelerated Electron Beam},
  author = {Cole, J. M. and Behm, K. T. and Gerstmayr, E. and Blackburn, T. G. and Wood, J. C. and Baird, C. D. and Duff, M. J. and Harvey, C. and Ilderton, A. and Joglekar, A. S. and others},
  journal = {Phys. Rev. X},
  volume = {8},
  issue = {1},
  pages = {011020},
  numpages = {11},
  year = {2018},
  month = {Feb},
  publisher = {American Physical Society},
  doi = {10.1103/PhysRevX.8.011020},
  url = {https://link.aps.org/doi/10.1103/PhysRevX.8.011020}
}

@article{poder_prx2018,
  title = {Experimental Signatures of the Quantum Nature of Radiation Reaction in the Field of an Ultraintense Laser},
 author = {Poder, K. and Tamburini, M. and Sarri, G. and Di Piazza, A. and Kuschel, S. and Baird, C. D. and Behm, K. and Bohlen, S. and Cole, J. M. and Corvan, D. J. and others},
  journal = {Phys. Rev. X},
  volume = {8},
  issue = {3},
  pages = {031004},
  numpages = {11},
  year = {2018},
  month = {Jul},
  publisher = {American Physical Society},
  doi = {10.1103/PhysRevX.8.031004},
  url = {https://link.aps.org/doi/10.1103/PhysRevX.8.031004}
}

@article{sauter1931,
  title={{\"U}ber das {V}erhalten eines {E}lektrons im homogenen elektrischen {F}eld nach der relativistischen {T}heorie {D}iracs},
  author={Sauter, Fritz},
  journal={Z. Phys.},
  volume={69},
  number={11},
  pages={742--764},
  year={1931},
  doi={10.1007/BF01339461},
  publisher={Springer}
}

@article{gonoskov_pre2015,
  title = {Extended particle-in-cell schemes for physics in ultrastrong laser fields: Review and developments},
  author = {Gonoskov, A. and Bastrakov, S. and Efimenko, E. and Ilderton, A. and Marklund, M. and Meyerov, I. and Muraviev, A. and Sergeev, A. and Surmin, I. and Wallin, E.},
  journal = {Phys. Rev. E},
  volume = {92},
  issue = {2},
  pages = {023305},
  numpages = {18},
  year = {2015},
  month = {Aug},
  publisher = {American Physical Society},
  doi = {10.1103/PhysRevE.92.023305},
  url = {https://link.aps.org/doi/10.1103/PhysRevE.92.023305}
}

@article{fedotov_physrep2023,
  title={Advances in QED with intense background fields},
  author={Fedotov, A and Ilderton, A and Karbstein, F and King, Ben and Seipt, D and Taya, H and Torgrimsson, Greger},
  journal={Phys. Rep.},
  volume={1010},
  pages={1--138},
  year={2023},
  publisher={Elsevier}
}

@article{gonoskov_rmp2022,
  title = {Charged particle motion and radiation in strong electromagnetic fields},
  author = {Gonoskov, A. and Blackburn, T. G. and Marklund, M. and Bulanov, S. S.},
  journal = {Rev. Mod. Phys.},
  volume = {94},
  issue = {4},
  pages = {045001},
  numpages = {63},
  year = {2022},
  month = {Oct},
  publisher = {American Physical Society},
  doi = {10.1103/RevModPhys.94.045001},
  url = {https://link.aps.org/doi/10.1103/RevModPhys.94.045001}
}

@inbook{schwinger1945,
  author={Schwinger, J.},
  year= 2000, 
  chapter={On radiation by electrons in a betatron, {R}eport {N}o. {LBNL}--39088}, 
  editor = {Kimball A. Milton}, 
  title= {A {Q}uantum {L}egacy: {S}eminal {P}apers of {J}ulian {S}chwinger}, 
  publisher= {World Scientific},
  address= {Singapore}, 
}

@article{michel_prl1982,
  title = {Intense Coherent Submillimeter Radiation in Electron Storage Rings},
  author = {Michel, F. Curtis},
  journal = {Phys. Rev. Lett.},
  volume = {48},
  issue = {9},
  pages = {580--583},
  numpages = {0},
  year = {1982},
  month = {Mar},
  publisher = {American Physical Society},
  doi = {10.1103/PhysRevLett.48.580},
  url = {https://link.aps.org/doi/10.1103/PhysRevLett.48.580}
}

@article{hirschmugl_pra1991,
  title = {Multiparticle coherence calculations for synchrotron-radiation emission},
  author = {Hirschmugl, Carol J. and Sagurton, Michael and Williams, Gwyn P.},
  journal = {Phys. Rev. A},
  volume = {44},
  issue = {2},
  pages = {1316--1320},
  numpages = {0},
  year = {1991},
  month = {Jul},
  publisher = {American Physical Society},
  doi = {10.1103/PhysRevA.44.1316},
  url = {https://link.aps.org/doi/10.1103/PhysRevA.44.1316}
}

@article{chen_ppcf2010,
  title={Radiation reaction effects on ion acceleration in laser foil interaction},
  author={Chen, Min and Pukhov, Alexander and Yu, Tong-Pu and Sheng, Zheng-Ming},
  journal={Plasma Physics and Controlled Fusion},
  volume={53},
  number={1},
  pages={014004},
  year={2010},
  publisher={IOP Publishing}
}

@article{gelfer_scirep2018,
  title={Unexpected impact of radiation friction: enhancing production of longitudinal plasma waves},
  author={Gelfer, Evgeny and Elkina, Nina and Fedotov, Alexander},
  journal={Sci. Rep.},
  volume={8},
  number={1},
  pages={6478},
  year={2018},
  publisher={Nature Publishing Group}
}

@article{gelfer_ppcf2018,
  title={Theory and simulations of radiation friction induced enhancement of laser-driven longitudinal fields},
  author={Gelfer, EG and Fedotov, AM and Weber, S},
  journal={Plasma Phys. Controlled Fusion},
  volume={60},
  number={6},
  pages={064005},
  year={2018},
  publisher={IOP Publishing}
}

@article{liseykina_njp2016,
  title={Inverse {F}araday effect driven by radiation friction},
  author={Liseykina, TV and Popruzhenko, SV and Macchi, A},
  journal={New J. Phys.},
  volume={18},
  number={7},
  pages={072001},
  year={2016},
  publisher={IOP Publishing}
}

@article{gelfer_njp2021,
  title={Radiation induced acceleration of ions in a laser irradiated transparent foil},
  author={Gelfer, EG and Fedotov, AM and Weber, S},
  journal={New J. Phys.},
  volume={23},
  number={9},
  pages={095002},
  year={2021},
  publisher={IOP Publishing}
}

@article{esarey_pre1993,
  title = {Nonlinear {Thomson} scattering of intense laser pulses from beams and plasmas},
  author = {Esarey, Eric and Ride, Sally K. and Sprangle, Phillip},
  journal = {Phys. Rev. E},
  volume = {48},
  issue = {4},
  pages = {3003--3021},
  numpages = {0},
  year = {1993},
  month = {Oct},
  publisher = {American Physical Society},
  doi = {10.1103/PhysRevE.48.3003},
  url = {https://link.aps.org/doi/10.1103/PhysRevE.48.3003}
}

@article{dirac_prs1938,
  title={Classical theory of radiating electrons},
  author={Dirac, Paul Adrien Maurice},
  journal={Proc. R. Soc. A},
  volume={167},
  number={929},
  pages={148--169},
  year={1938},
  publisher={The Royal Society London}
}

@book{jackson_book,
  title={Classical electrodynamics},
  author={Jackson, John David},
  year={2021},
  publisher={John Wiley \& Sons}
}

@article{eliezer_prs1948,
  title={On the classical theory of particles},
  author={Eliezer, C Jayaratnam},
  journal={Proceedings of the Royal Society of London. Series A. Mathematical and Physical Sciences},
  volume={194},
  number={1039},
  pages={543--555},
  year={1948},
  publisher={The Royal Society London}
}

@article{ford_pla1991,
  title={Radiation reaction in electrodynamics and the elimination of runaway solutions},
  author={Ford, GW and O'Connell, RF},
  journal={Phys. Lett. A},
  volume={157},
  number={4-5},
  pages={217--220},
  year={1991},
  publisher={Elsevier}
}

@article{burton_contphys2014,
  title={Aspects of electromagnetic radiation reaction in strong fields},
  author={Burton, David A and Noble, Adam},
  journal={Contemp. Phys.},
  volume={55},
  number={2},
  pages={110--121},
  year={2014},
  publisher={Taylor \& Francis}
}

@article{koga_pop2005,
    author = {Koga, James and Esirkepov, Timur Zh. and Bulanov, Sergei V.},
    title = "{Nonlinear {T}homson scattering in the strong radiation damping regime}",
    journal = {Physics of Plasmas},
    volume = {12},
    number = {9},
    pages = {093106},
    year = {2005},
    month = {09},
    issn = {1070-664X},
    doi = {10.1063/1.2013067},
    url = {https://doi.org/10.1063/1.2013067},
}

@article{dipiazza_lmp2008,
  title={Exact solution of the {L}andau--{L}ifshitz equation in a plane wave},
  author={Piazza, A Di},
  journal={Lett. Math. Phys.},
  volume={83},
  pages={305--313},
  year={2008},
  publisher={Springer}
}

@article{voronin_jetp1965,
  title={The pressure of an intense plane wave on a free charge and on a charge in a magnetic field},
  author={Voronin, BS and Kolomenskii, AA},
  journal={Sov. Phys. JETP},
  volume={20},
  pages={1027},
  year={1965}
}

@article{zeldovich_ufn1975,
  title={Interaction of free electrons with electromagnetic radiation},
  author={Zel'Dovich, Ya B},
  journal={Sov. Phys. Usp.},
  volume={18},
  number={2},
  pages={79},
  year={1975},
  publisher={IOP Publishing}
}

@article{popruzhenko_njp2019,
  title={Efficiency of radiation friction losses in laser-driven ‘hole boring’of dense targets},
  author={Popruzhenko, SV and Liseykina, TV and Macchi, A},
  journal={New J. Phys.},
  volume={21},
  number={3},
  pages={033009},
  year={2019},
  publisher={IOP Publishing}
}

@article{gelfer_prr2024,
  title = {Coherent radiation of an electron bunch colliding with an intense laser pulse},
  author = {Gelfer, E. G. and Fedotov, A. M. and Klimo, O. and Weber, S.},
  journal = {Phys. Rev. Res.},
  volume = {6},
  issue = {3},
  pages = {L032013},
  numpages = {7},
  year = {2024},
  month = {Jul},
  publisher = {American Physical Society},
  doi = {10.1103/PhysRevResearch.6.L032013},
  url = {https://link.aps.org/doi/10.1103/PhysRevResearch.6.L032013}
}

@article{gelfer_mre2024,
    author = {Gelfer, E. G. and Fedotov, A. M. and Klimo, O. and Weber, S.},
    title = "{Collective coherent emission of electrons in strong laser fields and perspective for hard x-ray lasers}",
    journal = {Matter Radiat. Extremes},
    volume = {9},
    number = {2},
    pages = {024201},
    year = {2024},
    month = {02},
    issn = {2468-2047},
    doi = {10.1063/5.0174508},
    url = {https://doi.org/10.1063/5.0174508},
}

@article{hartemann_pre2000,
  title={Stochastic electron gas theory of coherence in laser-driven synchrotron radiation},
  author={Hartemann, FV},
  journal={Phys. Rev. E},
  volume={61},
  number={1},
  pages={972},
  year={2000},
  publisher={APS}
}

@article{schiff_rsi1946,
    author = {Schiff, L. I.},
    title = "{Production of Particle Energies beyond 200 {Mev}}",
    journal = {Rev. Sci. Instrum.},
    volume = {17},
    number = {1},
    pages = {6-14},
    year = {1946},
    month = {01},
    issn = {0034-6748},
    doi = {10.1063/1.1770395},
    url = {https://doi.org/10.1063/1.1770395},
}

@article{thomas_prstab2010,
  title = {Algorithm for calculating spectral intensity due to charged particles in arbitrary motion},
  author = {Thomas, A. G. R.},
  journal = {Phys. Rev. ST Accel. Beams},
  volume = {13},
  issue = {2},
  pages = {020702},
  numpages = {11},
  year = {2010},
  month = {Feb},
  publisher = {American Physical Society},
  doi = {10.1103/PhysRevSTAB.13.020702},
  url = {https://link.aps.org/doi/10.1103/PhysRevSTAB.13.020702}
}

@article{kharin_pra2016,
  title = {Temporal laser-pulse-shape effects in nonlinear {T}homson scattering},
  author = {Kharin, V. Yu. and Seipt, D. and Rykovanov, S. G.},
  journal = {Phys. Rev. A},
  volume = {93},
  issue = {6},
  pages = {063801},
  numpages = {8},
  year = {2016},
  month = {Jun},
  publisher = {American Physical Society},
  doi = {10.1103/PhysRevA.93.063801},
  url = {https://link.aps.org/doi/10.1103/PhysRevA.93.063801}
}

@article{seipt_lasphys2013,
doi = {10.1088/1054-660X/23/7/075301},
url = {https://dx.doi.org/10.1088/1054-660X/23/7/075301},
year = {2013},
month = {may},
publisher = {IOP Publishing},
volume = {23},
number = {7},
pages = {075301},
author = {D Seipt and B Kämpfer},
title = {Nonlinear {C}ompton scattering of ultrahigh-intensity laser pulses},
journal = {Laser Phys.},
}

@article{boca_pra2009,
  title = {Nonlinear {Compton} scattering with a laser pulse},
  author = {Boca, Madalina and Florescu, Viorica},
  journal = {Phys. Rev. A},
  volume = {80},
  issue = {5},
  pages = {053403},
  numpages = {14},
  year = {2009},
  month = {Nov},
  publisher = {American Physical Society},
  doi = {10.1103/PhysRevA.80.053403},
  url = {https://link.aps.org/doi/10.1103/PhysRevA.80.053403}
}

@article{seipt_pra2011,
  title = {Nonlinear {Compton} scattering of ultrashort intense laser pulses},
  author = {Seipt, D. and K\"ampfer, B.},
  journal = {Phys. Rev. A},
  volume = {83},
  issue = {2},
  pages = {022101},
  numpages = {12},
  year = {2011},
  month = {Feb},
  publisher = {American Physical Society},
  doi = {10.1103/PhysRevA.83.022101},
  url = {https://link.aps.org/doi/10.1103/PhysRevA.83.022101}
}

@article{seipt_pra2013,
  title = {Asymmetries of azimuthal photon distributions in nonlinear {C}ompton scattering in ultrashort intense laser pulses},
  author = {Seipt, D. and K\"ampfer, B.},
  journal = {Phys. Rev. A},
  volume = {88},
  issue = {1},
  pages = {012127},
  numpages = {13},
  year = {2013},
  month = {Jul},
  publisher = {American Physical Society},
  doi = {10.1103/PhysRevA.88.012127},
  url = {https://link.aps.org/doi/10.1103/PhysRevA.88.012127}
}

@article{hartemann_pre1996,
  title={Spectral analysis of the nonlinear relativistic {Doppler} shift in ultrahigh intensity {Compton} scattering},
  author={Hartemann, FV and Troha, AL and Luhmann Jr, NC and Toffano, Zeno},
  journal={Phys. Rev. E},
  volume={54},
  number={3},
  pages={2956},
  year={1996},
  publisher={APS}
}

@article{salehi_prx2021,
  title = {Laser-Accelerated, Low-Divergence 15-{MeV} Quasimonoenergetic Electron Bunches at 1 {kHz}},
  author = {Salehi, F. and Le, M. and Railing, L. and Kolesik, M. and Milchberg, H. M.},
  journal = {Phys. Rev. X},
  volume = {11},
  issue = {2},
  pages = {021055},
  numpages = {12},
  year = {2021},
  month = {Jun},
  publisher = {American Physical Society},
  doi = {10.1103/PhysRevX.11.021055},
  url = {https://link.aps.org/doi/10.1103/PhysRevX.11.021055}
}

@article{storey_prstab2024,
  title = {Wakefield generation in hydrogen and lithium plasmas at {FACET-II}: Diagnostics and first beam-plasma interaction results},
  author = {Storey, D. and Zhang, C. and San Miguel Claveria, P. and Cao, G. J. and Adli, E. and Alsberg, L. and Ariniello, R. and Clarke, C. and Corde, S. and Dalichaouch, T. N. and others},
  journal = {Phys. Rev. Accel. Beams},
  volume = {27},
  issue = {5},
  pages = {051302},
  numpages = {15},
  year = {2024},
  month = {May},
  publisher = {American Physical Society},
  doi = {10.1103/PhysRevAccelBeams.27.051302},
  url = {https://link.aps.org/doi/10.1103/PhysRevAccelBeams.27.051302}
}

@article{yakimenko_prl2019,
  title = {Prospect of Studying Nonperturbative {QED} with Beam-Beam Collisions},
  author = {Yakimenko, V. and Meuren, S. and Del Gaudio, F. and Baumann, C. and Fedotov, A. and Fiuza, F. and Grismayer, T. and Hogan, M. J. and Pukhov, A. and Silva, L. O. and White, G.},
  journal = {Phys. Rev. Lett.},
  volume = {122},
  issue = {19},
  pages = {190404},
  numpages = {7},
  year = {2019},
  month = {May},
  publisher = {American Physical Society},
  doi = {10.1103/PhysRevLett.122.190404},
  url = {https://link.aps.org/doi/10.1103/PhysRevLett.122.190404}
}

@article{chang_prappl2023,
  title = {Reduction of the electron-beam divergence of laser wakefield accelerators by integrated plasma lenses},
  author = {Chang, Y.-Y. and Cabada\ifmmode \breve{g}\else \u{g}\fi{}, J. Couperus and Debus, A. and Ghaith, A. and LaBerge, M. and Pausch, R. and Sch\"obel, S. and Ufer, P. and Schramm, U. and Irman, A.},
  journal = {Phys. Rev. Appl.},
  volume = {20},
  issue = {6},
  pages = {L061001},
  numpages = {6},
  year = {2023},
  month = {Dec},
  publisher = {American Physical Society},
  doi = {10.1103/PhysRevApplied.20.L061001},
  url = {https://link.aps.org/doi/10.1103/PhysRevApplied.20.L061001}
}

@online{facet2,
    howpublished = {\url{https://facet-ii.slac.stanford.edu/facility/beam-parameters}},
    title = {{FACET II} beam parameters}
}

@article{tamburini_nimpra2011,
  title={Radiation reaction effects on electron nonlinear dynamics and ion acceleration in laser--solid interaction},
  author={Tamburini, Matteo and Pegoraro, Francesco and Di Piazza, Antonino and Keitel, Christoph H and Liseykina, Tatyana V and Macchi, Andrea},
  journal={Nucl. Instrum. Methods Phys. Res., Sect. A},
  volume={653},
  number={1},
  pages={181--185},
  year={2011},
  publisher={Elsevier}
}

@article{neitz_prl2013,
  title={Stochasticity effects in quantum radiation reaction},
  author={Neitz, Norman and Di Piazza, Antonino},
  journal={Phys. Rev. Lett.},
  volume={111},
  number={5},
  pages={054802},
  year={2013},
  publisher={APS}
}

@article{tamburini_pre2012,
  title={Radiation-pressure-dominant acceleration: Polarization and radiation reaction effects and energy increase in three-dimensional simulations},
  author={Tamburini, M and Liseykina, TV and Pegoraro, Francesco and Macchi, A},
  journal={Phys. Rev. E},
  volume={85},
  number={1},
  pages={016407},
  year={2012},
  publisher={APS}
}

@article{tamburini_njp2010,
  title={Radiation reaction effects on radiation pressure acceleration},
  author={Tamburini, M and Pegoraro, Francesco and Di Piazza, A and Keitel, Ch H and Macchi, A},
  journal={New J. Phys.},
  volume={12},
  number={12},
  pages={123005},
  year={2010},
  publisher={IOP Publishing}
}

@article{capdessus_pre2015,
  title={Influence of radiation reaction force on ultraintense laser-driven ion acceleration},
  author={Capdessus, R and McKenna, Paul},
  journal={Phys. Rev. E},
  volume={91},
  number={5},
  pages={053105},
  year={2015},
  publisher={APS}
}

@article{dipiazza_prl2010,
  title={Quantum radiation reaction effects in multiphoton {Compton} scattering},
  author={Di Piazza, A and Hatsagortsyan, KZ and Keitel, Christoph H},
  journal={Phys. Rev. Lett.},
  volume={105},
  number={22},
  pages={220403},
  year={2010},
  publisher={APS}
}

@article{thomas_prx2012,
  title={Strong radiation-damping effects in a gamma-ray source generated by the interaction of a high-intensity laser with a wakefield-accelerated electron beam},
  author={Thomas, AGR and Ridgers, CP and Bulanov, SS and Griffin, BJ and Mangles, SPD},
  journal={Phys. Rev. X},
  volume={2},
  number={4},
  pages={041004},
  year={2012},
  publisher={APS}
}

@article{vieira_natphys2021,
  title={Generalized superradiance for producing broadband coherent radiation with transversely modulated arbitrarily diluted bunches},
  author={Vieira, J and Pardal, Miguel and Mendon{\c{c}}a, JT and Fonseca, RA},
  journal={Nat. Phys.},
  volume={17},
  number={1},
  pages={99--104},
  year={2021},
  publisher={Nature Publishing Group UK London}
}

@article{quin_ppcf2025,
  title={Coherent frequency combs from electrons colliding with a laser pulse},
  author={Quin, Michael J and Di Piazza, Antonino and Tamburini, Matteo},
  journal={Plasma Physics and Controlled Fusion},
  volume={67},
  number={5},
  pages={055008},
  year={2025},
  publisher={IOP Publishing}
}

@article{quin_prr2025,
  title = {Effect of interparticle fields and radiation reaction on beam dynamics},
  author = {Quin, Michael J. and Di Piazza, Antonino and Keitel, Christoph H. and Tamburini, Matteo},
  journal = {Phys. Rev. Res.},
  volume = {7},
  issue = {2},
  pages = {023210},
  numpages = {12},
  year = {2025},
  month = {Jun},
  publisher = {American Physical Society},
  doi = {10.1103/PhysRevResearch.7.023210},
  url = {https://link.aps.org/doi/10.1103/PhysRevResearch.7.023210}
}

@article{malaca_natphot2024,
  title={Coherence and superradiance from a plasma-based quasiparticle accelerator},
  author={Malaca, B and Pardal, M and Ramsey, D and Pierce, JR and Weichman, K and Andriyash, IA and Mori, WB and Palastro, JP and Fonseca, RA and Vieira, J},
  journal={Nat. Phot.},
  volume={18},
  number={1},
  pages={39--45},
  year={2024},
  publisher={Nature Publishing Group UK London},
  url={https://www.nature.com/articles/s41566-023-01311-z#citeas}
}

@article{prl_arxiv,
  title={Coherently enhanced radiation friction in laser--plasma collisions},
  author={Gelfer, EG and Fedotov, AM and Malakhov, MP and Klimo, O and Weber, S},
  journal={arXiv preprint arXiv:2502.20290},
  year={2025}
}

@article{los_natcomm2026,
  title={Observation of quantum effects on radiation reaction in strong fields},
  author={Los, Eva E and Gerstmayr, Elias and Arran, Christopher and Streeter, Matthew JV and Colgan, Cary and Cobo, Claudia C and Kettle, Brendan and Blackburn, Thomas G and Bourgeois, Nicolas and Calvin, Luke and others},
  journal={Nature Communications},
  year={2026},
  publisher={Nature Publishing Group UK London}
}

@article{malakhov_prep,
  title={Analytical calculation of the spectrum of nonlinear Compton scattering},
  author={Malakhov, M.P. and Benahmed, Th. and Gelfer, E.G. and Fedotov, A.M. and Klimo O. and  Weber, S. and Rykovanov, S.G.},
  journal={in preparation},
  year={2026},
}

@article{blackman_commphys2022,
  title={Electron acceleration from transparent targets irradiated by ultra-intense helical laser beams},
  author={Blackman, David R and Shi, Yin and Klein, Sallee R and Cernaianu, Mihail and Doria, Domenico and Ghenuche, Petru and Arefiev, Alexey},
  journal={Communications Physics},
  volume={5},
  number={1},
  pages={116},
  year={2022},
  publisher={Nature Publishing Group UK London}
}

@article{shi_hplse2022,
  title={Electron pulse train accelerated by a linearly polarized Laguerre--Gaussian laser beam},
  author={Shi, Yin and Blackman, David R and Zhu, Ping and Arefiev, Alexey},
  journal={High Power Laser Science and Engineering},
  volume={10},
  pages={e45},
  year={2022},
  publisher={Cambridge University Press}
}

@article{shi_ppcf2021,
  title={Electron acceleration using twisted laser wavefronts},
  author={Shi, Yin and R Blackman, David and Arefiev, Alexey},
  journal={Plasma Physics and Controlled Fusion},
  volume={63},
  number={12},
  pages={125032},
  year={2021},
  publisher={IOP Publishing}
}

@article{shi_prl2021,
  title={Generation of ultrarelativistic monoenergetic electron bunches via a synergistic interaction of longitudinal electric and magnetic fields of a twisted laser},
  author={Shi, Yin and Blackman, David and Stutman, Dan and Arefiev, Alexey},
  journal={Physical Review Letters},
  volume={126},
  number={23},
  pages={234801},
  year={2021},
  publisher={APS}
}

\end{document}